\documentclass[
aps,
physrev,
amsmath,
physrev
]{revtex4-2}

\usepackage{blindtext}
\usepackage{graphicx}
\usepackage{dcolumn}
\usepackage{bm}
\usepackage[utf8]{inputenc}
\usepackage[T1]{fontenc}
\usepackage{mathptmx}
\usepackage{etoolbox}
\usepackage{upgreek} 
\usepackage[normalem]{ulem} 

\usepackage{amsmath}
\usepackage{amsthm}
\usepackage{amssymb}
\usepackage{esint}
\usepackage[linkcolor=blue,colorlinks=true]{hyperref}
\usepackage{multirow}

\usepackage[normalem]{ulem}

\makeatletter 

\newenvironment{newnumbering*}[1][Alph]
 {%
  \def\jr@counter{#1}%
  \jr@setup@numbering{#1}{0}%
 }
 {%
  \setcounter{jr@\jr@counter @equation}{\value{equation}}%
  \setcounter{equation}{\value{jr@equation@save}}%
  \ignorespacesafterend
 }

\newcounter{jr@equation@save}
\newcommand{\jr@setup@numbering}[2]{%
  \@ifundefined{c@jr@#1@equation}{\newcounter{jr@#1@equation}}{}%
  \setcounter{jr@equation@save}{\value{equation}}%
  \setcounter{equation}{%
    \ifnum#2>0
      \value{jr@#1@equation}%
    \else
      0%
    \fi
  }%
  \renewcommand{\theequation}{\csname#1\endcsname{equation}}%
}
\makeatother

\makeatletter
\def\@email#1#2{%
 \endgroup
 \patchcmd{\titleblock@produce}
  {\frontmatter@RRAPformat}
  {\frontmatter@RRAPformat{\produce@RRAP{*#1\href{mailto:#2}{#2}}}\frontmatter@RRAPformat}
  {}{}
}%
\makeatother

\begin{document}

\title[Variational hydrodynamics of the classical Yukawa one-component plasma]{Variational hydrodynamics of the classical Yukawa one-component plasma}
\author{Daniels Krimans}
\email{krimans@physik.uni-kiel.de}
\affiliation{Institute of Theoretical Physics and Astrophysics, Christian-Albrechts-Universität zu Kiel, Kiel, Germany}
\author{Hanno Kählert}
\affiliation{Institute of Theoretical Physics and Astrophysics, Christian-Albrechts-Universität zu Kiel, Kiel, Germany}

\date{June 28, 2025}

\begin{abstract} 
We consider a recently developed variational approach to the hydrodynamics of strongly coupled plasmas [D.~Krimans and S.~Putterman, Phys. Fluids \textbf{36}, 037131 (2024)] and extend it to the Yukawa one-component plasma. This approach generalizes the ordinary hydrodynamic equations to finite length scales by explicitly including terms that depend on the pair distribution function. After discussing the form of the Lagrangian, we derive equations of motion and explicit formulas for the momentum and energy conservation laws.
After demonstrating consistency with thermodynamics, we consider the simpler linear regime and the dispersion laws. By comparing the longitudinal speed of sound to existing numerical data, we find excellent agreement in the weak to moderate screening regimes, while discrepancies arise at strong screening.
The finite-wavelength behavior of the longitudinal dispersion relation also shows excellent agreement with simulations across a wide range of coupling and screening parameters, even when the wavelength is comparable to the average interparticle spacing. In addition to the linear regime, our variational approach has potential for application to nonlinear problems and other physical systems.
\end{abstract}

\maketitle

\section{Introduction}

Strongly coupled plasmas constitute an intriguing and challenging regime of plasma physics in which particle interactions and correlations dominate the dynamics.
Such plasmas have been observed experimentally and explored theoretically across a wide range of systems, including ultracold neutral plasmas \cite{KILLIAN200777, Lyon_2017, PhysRevLett.105.065004, 10.1063/1.3694654, PhysRevE.105.045201, 10.1063/5.0136369}, dusty plasmas \cite{RevModPhys.81.1353, Bonitz_2010, BANDYOPADHYAY2007491, PhysRevLett.103.115002}, warm dense matter \cite{PhysRevLett.98.065002,PhysRevE.90.033107, PhysRevB.99.165103, PhysRevResearch.6.L022029}, or during laser breakdown in dense gases \cite{Bataller2019DynamicsOS}. 
Strong coupling physics also plays an important role in molten metals \cite{PhysRevB.77.174203, 10.1063/1.4794661, Hosokawa_2015} and molten salts \cite{DEMMEL200598, Demmel_2021}. 
The availability of experimental data on various linear and nonlinear phenomena motivates the development of a theoretical framework capable of describing their time evolution.
\par 

Even for the simplest hydrogen plasma, incorporating strong coupling effects into a rigorous theoretical description remains challenging \cite{10.1063/5.0219405}, and numerous theoretical tools exist, each suited to particular length and timescales or specific phenomena.
A common source of theoretical difficulty in strongly coupled systems is the emergence of particle correlations, evident in the pair distribution function, which gives the probability of finding two particles separated by a given distance.
Since typical experimental systems are macroscopic and thus consist of a large number of particles, the most precise direct approaches, such as particle-based simulations, require extensive computational resources, often making the calculations impractical.
This limitation motivates the search for approximate, yet accurate descriptions.
In this work, we consider a hydrodynamic approach to plasma dynamics, leading to equations that are physically clear, independent of the exact number of particles, and computationally efficient even over long timescales.
However, typical hydrodynamic theories do not include the strong coupling effects important for these plasmas, so our goal is to appropriately extend them.
\par 

To formulate hydrodynamic equations suitable for strongly coupled plasmas we employ a recently proposed variational framework \cite{10.1063/5.0194352}, developed for arbitrary interaction potentials and explicitly incorporating the pair distribution function. 
Variational principles allow the dynamics to be formulated in terms of a single Lagrangian in which each term has a clear physical meaning, and guarantee the self-consistency of the resulting equations of motion with the associated conservation laws \cite{landau1976classical, landau1980fields, herivel_1955, doi:10.1098/rspa.1968.0103}.
For example, explicit inclusion of the pair distribution function in the Lagrangian yields interaction terms that capture correlation effects without compromising momentum or energy conservation.
This variational principle was originally applied to the simplest model of a strongly coupled plasma, the one‑component plasma (OCP) \cite{10.1063/5.0194352}.
In the linear regime, it was shown to agree with the molecular dynamics (MD) data across the entire coupling range\cite{10.1063/1.3679586, 10.1063/5.0229805}, even at wavelengths comparable to the interparticle spacing.
\par

Encouraged by the success of the OCP, we apply the same framework to the more complex Yukawa one‑component plasma (YOCP), in which a single species of particles moves in a neutralizing background that provides screening effects.
This model has been used to describe ultracold neutral plasmas \cite{KILLIAN200777, Lyon_2017}, dusty plasmas \cite{RevModPhys.81.1353, Bonitz_2010}, and molten metals \cite{10.1063/1.5088141}, and is particularly accurate when one species is much lighter than the other, for example, in the case of ions and electrons, allowing for a separation of timescales in their dynamics.
To include screening, the interaction between moving particles is modeled using the screened Coulomb potential $\phi(r) = q^2 e^{-r/\lambda}/(4 \pi \varepsilon_0 r)$, where $q$ is the particle charge, $\lambda$ is the screening length, and $r$ is the interparticle separation.
This simplified model is useful because only one species is dynamic, and its thermodynamic properties are well characterized \cite{Hansen_McDonald_2005, PhysRevE.90.053101, PhysRevE.56.4671, Caillol_2000}, which are required as input for our model.
Simple fits for the energy and pressure are available \cite{PhysRevE.91.023108, Khrapak_2016} in terms of the dimensionless coupling parameter $\Gamma = q^2/(4\pi \varepsilon_0 a k_B T)$ and the dimensionless screening parameter $\kappa = a/\lambda$. Here, $a=[3/(4\pi n)]^{1/3}$ is the Wigner-Seitz radius, $n$ is the density, and $T$ is the temperature.
We focus on the regime $\Gamma > 1$, where interactions play a significant role.
\par 

The goal of this paper is to demonstrate that the variational framework can extend conventional hydrodynamic equations for the YOCP in the strongly coupled regime to small length scales comparable to the interparticle spacing.
We validate the approach by demonstrating consistency with thermodynamics in thermal equilibrium and by computing dispersion relations that clearly exhibit correlational effects.
As in the OCP case, the dispersion laws require the pair distribution function and its derivatives. 
Because we are not aware of a simple yet accurate fit for this function in the literature, we approximate it with a step function model that has already proved successful for the OCP \cite{10.1063/5.0194352} and within the quasi-localized charge approximation (QLCA) \cite{10.1063/1.4942169, 10.1063/1.5088141}.
\par

We examine the longitudinal dispersion relation for a broad range of screening parameters and coupling strengths and demonstrate very good agreement of our results with MD simulations, both our own and previously published data \cite{PhysRevE.100.063206}. 
We also compare with several well-known theoretical approaches, including the QLCA~\cite{PhysRevA.41.5516, 10.1063/1.873814, Donkó_2008}, the extended QLCA \cite{PhysRevE.79.046412, PhysRevE.87.043102, 10.1063/1.4965903, https://doi.org/10.1002/ctpp.202400018}, and Euler hydrodynamics with a mean-field term \cite{Khrapak_2016, 10.1063/1.2759881, RAO1990543, PhysRevE.91.033110}. 
Additional theoretical approaches can be found in the literature~\cite{arkhipov2017prl, arkhipov2020pre, PhysRevE.102.033207, tolias2021pop}.
Specifically, for the long-wavelength speed of sound, we find excellent agreement with MD data in the weak and moderate screening regimes.
For larger screening values the speed of sound is overestimated, whereas Euler hydrodynamics with a mean-field term remains highly accurate.
For finite wavelengths, the dispersion we obtain matches the MD data across a broad range of coupling and screening parameters.
Considering the full range of coupling strengths, from the weakly coupled limit up to the melting point, our theory provides a clear improvement over other analytic approaches that we tested.
\par 

In Section~\ref{section:variational_principle_and_conservation_laws} we summarize the variational principle and derive the equations of motion together with the associated conservation laws.
Section \ref{section:equilibrium_solutions} analyzes solutions in thermal equilibrium, while Section \ref{section:linearized_solutions} derives the longitudinal and transverse dispersion relations.
In Section \ref{section:thermodynamics_and_step_function_approximation}, we review the thermodynamic properties of the YOCP and introduce the step function approximation for the pair distribution function.
Section~\ref{section:results_and_comparison} examines the longitudinal dispersion relation over a broad range of coupling and screening parameters and compares our predictions to MD simulations and other analytic approaches.
Finally, Section~\ref{section:conclusions} summarizes the results and outlines possible improvements and future applications.

\section{Variational principle and conservation laws}
\label{section:variational_principle_and_conservation_laws}

To construct a variational principle for the YOCP, we use the variational approach outlined in Ref.~\onlinecite{10.1063/5.0194352}, originally developed for the OCP.
Unlike the OCP treatment, we formulate the principle solely for the mobile charged particles and do not explicitly include the particles that constitute the neutralizing background.
Although the background is nonuniform due to the assumption of the Boltzmann distribution, its influence is already captured by the effective screened potential between the mobile particles \cite{10.1063/1.467954}.
In the OCP case, the background had to be included to ensure convergence of the total energy.
\par

The variational principle for a system of $N$ point particles with trajectories $\vec{x}_a(t)$, where $a = 1, 2, ..., N$ and $t$ is time, interacting through a pairwise interaction potential $\phi$, is expressed by the Lagrangian~\cite{landau1976classical}
\begin{equation}
\label{eq:N_particle_Lagrangian}
L = \sum \limits_{a=1}^N \frac{1}{2}m \left( \frac{\mathrm{d}\vec{x}_a}{\mathrm{d}t} \right)^2 - \frac{1}{2} \sum \limits_{a=1}^N \sum \limits_{\substack{a'=1 \\ a'\neq a}}^N \phi \left( |\vec{x}_a - \vec{x}_{a'}| \right),
\end{equation}
where the first sum is the total kinetic energy, and the second is the total potential energy.
Motivated by this $N$-particle Lagrangian, our goal is to construct an analogous, averaged Lagrangian that describes hydrodynamic motion for interactions of arbitrary form and strength.
\par 

As detailed in Ref.~\onlinecite{10.1063/5.0194352}, the variational principle is most naturally formulated in Lagrangian coordinates.
There, the hydrodynamic motion is described by the displacement field $\vec{x}$, which is the averaged macroscopic position at time $t$ of a particle whose position in some reference state is $\vec{a}$.
The reference state is usually taken as either the initial experimental configuration or a state of thermal equilibrium.
\par

The number density $n$ and the specific entropy $s$ in the laboratory frame are related to the time-independent number density $n^{\mathrm{ref}}$ and the specific entropy $s^{\mathrm{ref}}$ of the chosen reference state, as well as to the displacement field $\vec{x}$ \cite{herivel_1955, doi:10.1098/rspa.1968.0103}, by
\begin{equation}
	\label{eq:conservation_laws}
	\begin{gathered}
		n(\vec{x},t)=\frac{n^{\mathrm{ref}}}{\mathrm{det}(\partial \vec{x}/\partial \vec{a})}, \quad s(\vec{x},t)=s^{\mathrm{ref}}.
	\end{gathered}
\end{equation}
\par

The proposed Lagrangian for the YOCP is given by
\begin{equation}
	\label{eq:lagrangian}
	\begin{gathered}
		L = \int \left( \frac{1}{2}mn^{\mathrm{ref}} \left( \frac{\partial \vec{x}}{\partial t} \right)^2 - \frac{3}{2} n^{\mathrm{ref}} k_B T(n, s) \right) \mathrm{d}\vec{a} \\
		- \frac{1}{2} \iint n^{\mathrm{ref}} (n^{\mathrm{ref}})^T \phi(|\vec{x}-\vec{x}^T|) g(n^{\mathrm{ref}}, T(n, s), |\vec{a}-\vec{a}'|) \mathrm{d}\vec{a} \mathrm{d}\vec{a}',
	\end{gathered}
\end{equation}
and the corresponding equations of motion for the displacement field $\vec{x}$ follow from the variational principle.
The first two terms represent the first sum in Eq.~\eqref{eq:N_particle_Lagrangian}, which is split into the macroscopic and the averaged microscopic kinetic energies of the moving particles, each of mass $m$ and at temperature $T$, where $k_B$ is the Boltzmann constant.
If only these two local terms are retained, the equations of motion reduce to the standard Euler hydrodynamic equations for an ideal gas \cite{herivel_1955, doi:10.1098/rspa.1968.0103}.
\par

To formulate a variational principle for the YOCP that properly accounts for particle interactions even at strong coupling, we include the third term, which is nonlocal because it depends on the screened interaction potential $\phi$ and the pair distribution function $g$.
The pair distribution function appears because the double sum in Eq.~\eqref{eq:N_particle_Lagrangian} is averaged with the two-particle density rather than the one-particle density.
The screening length, $\lambda$, of the interaction potential is treated as a constant and its value can be related to the properties of the nonuniform neutralizing background \cite{10.1063/1.467954}.
The specific form of the pair distribution function out of equilibrium, together with its dependence on various variables, was analyzed in detail in Ref.~\onlinecite{10.1063/5.0194352} for the OCP.
It showed that this particular choice yields the most accurate results.
Here and henceforth, the superscript $f^T$ indicates that the function $f$ is obtained by interchanging all occurrences of $\vec{a}$ and $\vec{a}'$. 
For example, $f^T(\vec{a})=f(\vec{a}')$ and $f(\vec{x}, \vec{x}^T) = f(\vec{x}^T, \vec{x})$.
\par

Following the strategy of Ref.~\onlinecite{10.1063/5.0194352}, variation of $L$ yields for each Cartesian component $i=1,2,3$ the equations of motion 
\begin{equation}
	\label{eq:equations_of_motion}
	\begin{gathered}
		m n^{\mathrm{ref}} \frac{\partial^2 x_i}{\partial t^2} = \sum \limits_{j=1}^3 \frac{\partial (\mathrm{det}(\partial \vec{x}/\partial \vec{a}))}{\partial (\partial_j x_i)} \frac{\partial}{\partial a_j} \left( \frac{\partial T}{\partial (1/n)} \bigg|_s \left\{ \frac{3}{2} k_B + \frac{1}{2} \int (n^{\mathrm{ref}})^T \phi \frac{\partial g}{\partial T} \bigg|_{n^{\mathrm{ref}}, |\vec{a}-\vec{a}'|} \mathrm{d}\vec{a}' \right\} \right) 
		\\
		-n^{\mathrm{ref}} \int (n^{\mathrm{ref}})^T \frac{\partial \phi}{\partial x_i}\bigg|_{\vec{x}^T} \left( \frac{g + g^T}{2} \right) \mathrm{d}\vec{a}'.
	\end{gathered}
\end{equation}
The left-hand side represents the macroscopic acceleration of the particles.
The sum on the right-hand side accounts for the force arising from density variations.  
If the nonlocal term inside the curly brackets is omitted, corresponding to neglecting interaction, the quantity under the spatial derivative reduces to the ideal gas pressure.
This force is also appropriately weighted by the pair distribution function to correctly capture the statistical effects, which are particularly important at strong coupling.
\par

As shown in Ref.~\onlinecite{10.1063/5.0194352}, one can transform from the reference variables $\vec{a}$ and $\vec{a}'$ to the more commonly used laboratory coordinates $\vec{x}$ and $\vec{x}^T$.
In this representation, Eq.~\eqref{eq:equations_of_motion} takes the form 
\begin{equation}
	\label{eq:equations_of_motion_Eulerian}
	\begin{gathered}
		mn \left( \frac{\partial \vec{v}}{\partial t} + \left( \vec{v}\cdot \vec{\nabla} \right)\vec{v} \right) = \vec{\nabla} \left( \frac{\partial T}{\partial (1/n)} \bigg|_s \left\{ \frac{3}{2} k_B + \frac{1}{2} \int n^T \phi \frac{\partial g}{\partial T} \bigg|_{n^{\mathrm{ref}}, |\vec{a}-\vec{a}'|} \mathrm{d}\vec{x}^T \right\} \right) 
		\\
		-n\int n^T (\vec{\nabla}\phi) \big|_{\vec{x}^T} \left( \frac{g + g^T}{2} \right) \mathrm{d}\vec{x}^T,
	\end{gathered}
\end{equation}
where $\vec{v}$ denotes the velocity field in the laboratory frame, and the reference variables are now given by the functions $\vec{a}(\vec{x}, t)$ and $\vec{a}' = \vec{a}(\vec{x}^T, t)$.
Even in these coordinates, the pair distribution function still depends on the reference variables $\vec{a}, \vec{a}'$ as well as on the reference number density $n^{\mathrm{ref}}(\vec{a}(\vec{x}, t))$.
For this reason, it is often more convenient to analyze the equations in the reference frame.
\par

We now briefly discuss the conservation laws for momentum and energy, with particular emphasis on energy conservation, which is essential for consistency with thermodynamics.
The conserved energy, derived from the variational principle, will later be computed under conditions of thermal equilibrium, and we will verify that it coincides with the thermodynamic energy of the YOCP. 
Our derivations follow the treatment in Ref.~\onlinecite{10.1063/5.0194352}.
\par

We first consider the momentum conservation law in each of the Cartesian directions $i = 1, 2, 3$, as given by
\begin{equation}
	\label{eq:momentum_conservation_law}
	\begin{gathered}
		\frac{ \partial}{\partial t} \left( m n^{\mathrm{ref}} \frac{\partial x_i}{\partial t} \right) + \sum \limits_{j=1}^3  \frac{\partial}{\partial a_j} \left( - \frac{\partial (\mathrm{det}(\partial \vec{x}/\partial \vec{a}))}{\partial (\partial_j x_i)}  \frac{\partial T}{\partial (1/n)} \bigg|_s \left\{ \frac{3}{2} k_B + \frac{1}{2} \int (n^{\mathrm{ref}})^T \phi \frac{\partial g}{\partial T} \bigg|_{n^{\mathrm{ref}}, |\vec{a}-\vec{a}'|} \mathrm{d}\vec{a}' \right\} \right) 
		\\
		= -	\int n^{\mathrm{ref}}(n^{\mathrm{ref}})^T \frac{\partial \phi}{\partial x_i}\bigg|_{\vec{x}^T} \left( \frac{g + g^T}{2} \right) \mathrm{d}\vec{a}'.
	\end{gathered}
\end{equation}
This result relies on the fact that the local part of the Lagrangian in Eq.~\eqref{eq:lagrangian} does not depend explicitly on $\vec{x}$, while the nonlocal part explicitly depends on $\vec{x}$ and $\vec{x}^T$ only through the translationally invariant combination $\vec{x}-\vec{x}^T$.
Integrating Eq.~\eqref{eq:momentum_conservation_law} defines the total momentum $\vec{P}$ that satisfies
\begin{equation}
	\label{eq:momentum_conservation_law_v2}
	\begin{gathered}
		\frac{\mathrm{d} \vec{P}}{\mathrm{d} t} = \frac{\mathrm{d}}{\mathrm{d} t} \left( \int m n^{\mathrm{ref}} \frac{\partial \vec{x}}{\partial t} \mathrm{d}\vec{a}\right) = \vec{0}.
	\end{gathered}
\end{equation}
\par

Because neither the local nor the nonlocal term in the Lagrangian depends explicitly on time, an energy conservation law also exists.
Explicit expressions for the energy density, the energy flux, and the nonlocal source term are given in Appendix~\ref{appendix:energy_law}.
Here we merely note that integrating the energy conservation law yields the conserved total energy $E$, as
\begin{equation}
	\label{eq:energy_conservation_law_v2}
	\begin{gathered}
		\frac{\mathrm{d} E}{\mathrm{d} t} = \frac{\mathrm{d}}{\mathrm{d} t} \left( \int \left\{ \frac{1}{2}mn^{\mathrm{ref}} \left( \frac{\partial \vec{x}}{\partial t} \right)^2 + \frac{3}{2} n^{\mathrm{ref}} k_B T \right\} \mathrm{d}\vec{a} + \frac{1}{2} \iint n^{\mathrm{ref}} (n^{\mathrm{ref}})^T \phi  \left( \frac{g + g^T}{2} \right) \mathrm{d}\vec{a} \mathrm{d}\vec{a}' \right) = 0.
	\end{gathered}
\end{equation}
\par

\section{Equilibrium solutions}
\label{section:equilibrium_solutions}

We now consider equilibrium solutions of the YOCP in thermal equilibrium.  
Specifically, we verify that the equations of motion in Eq.~\eqref{eq:equations_of_motion} are satisfied by the equilibrium configuration and that the expression for the conserved energy in Eq.~\eqref{eq:energy_conservation_law_v2} reproduces the thermodynamic energy.
\par

If the reference configuration is chosen to coincide with the positions of the particles in equilibrium, the absence of macroscopic motion implies $\vec{x}(\vec{a}, t) = \vec{a}$.
Furthermore, the reference number density and the specific entropy are uniform and time-independent, taking the values $n^{\mathrm{ref}} = n_0$ and $s^{\mathrm{ref}} = s_0$. 
Eq.~\eqref{eq:conservation_laws} then yields uniform and time-independent fields in the laboratory frame, namely $n = n_0$ and $s = s_0$.
\par 

Substituting these equilibrium fields into Eq.~\eqref{eq:equations_of_motion} shows that the left-hand side vanishes because $\vec{x}$ is time-independent. 
The sum on the right-hand side also vanishes, since the differentiated quantity is independent of $\vec{a}$.  
For the remaining integral, we change the variables to $\vec{r} = \vec{a}-\vec{a}'$ and note that $\phi(|\vec{r}|)$ depends only on $r=|\vec{r}|$, implying that $\partial \phi/\partial r_i$ is odd in $r_i$, and its integral over all space therefore vanishes.
Hence, the proposed equilibrium solutions indeed satisfy the equations of motion.
\par

Next, we evaluate the equilibrium energy $E_0$.  
In the thermodynamic limit, where the particle number $N \to \infty$ and the system volume $V \to \infty$ at fixed number density $n_0 = N/V$, the total energy diverges.
Therefore, we instead computed the energy per particle. 
Inserting the equilibrium fields into Eq.~\eqref{eq:energy_conservation_law_v2} gives
\begin{equation}
	\label{eq:equilibrium_energy}
	\begin{gathered}
		\frac{E_0}{N} = \frac{3}{2} k_B T_0 + \frac{n_0}{2} \int \phi(|\vec{r}|) g(n_0, T_0, |\vec{r}|) \mathrm{d}\vec{r},
	\end{gathered}
\end{equation}
where the subscript ``$0$'' denotes evaluation at equilibrium.
\par

Eq.~\eqref{eq:equilibrium_energy} coincides with the standard thermodynamic result for interacting systems \cite{Hansen_McDonald_2005}.
For the OCP, $E_0/N$ diverges unless the neutralizing background is included \cite{Teller_1966}.
This divergence arises because \cite{Teller_1966, Hansen_McDonald_2005} $g(n_0, T_0, |\vec{r}|) \to 1$ as $|\vec{r}| \to \infty$ and the three-dimensional Coulomb potential is not integrable.
By contrast, the screened potential in the YOCP yields a finite integral:
\begin{equation}
	\begin{gathered}
		\int \phi(|\vec{r}|) \mathrm{d}\vec{r} = \frac{q^2}{4 \pi \varepsilon_0} \int \frac{e^{-|\vec{r}|/\lambda}}{|\vec{r}|} \mathrm{d}\vec{r} = \frac{q^2 \lambda^2}{\varepsilon_0}.
	\end{gathered}
\end{equation}
\par 

This shows that $E_0/N$ converges for the moving particles alone and the background can be neglected, greatly simplifying the analysis.  
Nevertheless, a consistent treatment of the limit $\lambda \to \infty$, in which the YOCP approaches the OCP, requires the proper inclusion of the neutralizing background \cite{10.1063/1.467954}.
\par

\section{Linearized solutions and the dispersion laws}
\label{section:linearized_solutions}

We now consider the linearized equations of motion and derive both the longitudinal and transverse dispersion laws, following the strategy detailed for the OCP in Ref.~\onlinecite{10.1063/5.0194352}.  
As before, we choose the reference state to correspond to thermal equilibrium.
The displacement field is assumed to take the form
\begin{equation}
	\label{eq:linearized_x}
	\vec{x}(\vec{a}, t) = \vec{a} + \vec{\xi}(\vec{a}, t),
\end{equation}
where $\vec{\xi}$ denotes the first-order correction.
\par 

Using this ansatz, we expand the equations of motion from Eq.~\eqref{eq:equations_of_motion} to first order in $\vec{\xi}$.
We then perform a Fourier transform of $\vec{\xi}$ with respect to the spatial coordinates $\vec{a}$, defining the transformed displacement field $\vec{\Psi}(\vec{k}, t)$.
Both of these steps are presented in detail in Appendix~\ref{appendix:linearized_eom}.
Using the rotational invariance of $\phi$ and $g$, we align the coordinate system such that the wave vector $\vec{k}$ points in the $z$-direction.
This yields three decoupled equations $\partial^2 \Psi_x/\partial t^2 = - \omega_T^2(|\vec{k}|) \Psi_x$, $\partial^2 \Psi_y/\partial t^2 = - \omega_T^2(|\vec{k}|) \Psi_y$, and $\partial^2 \Psi_z/\partial t^2 = - \omega_L^2(|\vec{k}|) \Psi_z$, where $\omega_T(|\vec{k}|)$ and $\omega_L(|\vec{k}|)$ are the transverse and longitudinal dispersion relations, respectively, given by
\begin{equation}
	\label{eq:general_transverse_law}
	\begin{gathered}
		\omega_T^2(|\vec{k}|) = \frac{q^2 n_0}{m \varepsilon_0} \int \limits_0^{\infty} \frac{e^{-r/\lambda}}{r} g(n_0, T_0, r) \Bigg( \left( 1 + \frac{r}{\lambda} + \frac{r^2}{3\lambda^2} \right) \\
		\times \left( \frac{\sin(|\vec{k}|r)}{|\vec{k}|r} + \frac{3\cos(|\vec{k}|r)}{|\vec{k}|^2r^2} - \frac{3\sin(|\vec{k}|r)}{|\vec{k}|^3r^3} \right) + \frac{r^2}{3\lambda^2} \left( 1 - \frac{\sin(|\vec{k}|r)}{|\vec{k}|r} \right) \Bigg) \mathrm{d}r,
	\end{gathered}
\end{equation}
\begin{equation}
	\label{eq:general_longitudinal_law}
	\begin{gathered}
		\omega_L^2(|\vec{k}|) = -|\vec{k}|^2 \Bigg( \frac{\partial}{\partial n_0} \bigg|_{s_0} \left( \frac{3}{2} \frac{k_B}{m} \frac{\partial T_0}{\partial (1/n_0)} \bigg|_{s_0}  \right)
		+ \frac{q^2 n_0}{2 m \varepsilon_0} \frac{\partial^2 T_0}{\partial n_0 \partial (1/n_0)} \bigg|_{s_0} \int \limits_0^{\infty} re^{-r/\lambda} \frac{\partial g_0}{\partial T_0} \bigg|_{n_0, r} \mathrm{d}r \\
		+\frac{q^2 n_0}{2 m \varepsilon_0} \frac{\partial T_0}{\partial (1/n_0)} \bigg|_{s_0} \frac{\partial T_0}{\partial n_0} \bigg|_{s_0} \int \limits_0^{\infty} re^{-r/\lambda} \frac{\partial^2 g_0}{\partial T_0^2} \bigg|_{n_0, r} \mathrm{d}r \bigg) \\
		+ \frac{q^2}{m \varepsilon_0}\frac{\partial T_0}{\partial (1/n_0)} \bigg|_{s_0} \int \limits_0^{\infty} 
		\frac{e^{-r/\lambda}}{r} \left( 1 + \frac{r}{\lambda} \right) \frac{\partial g_0}{\partial T_0} \bigg|_{n_0, r} \left( \cos\left( |\vec{k}|r \right) - \frac{\sin\left(|\vec{k}|r \right)}{|\vec{k}|r } \right) \mathrm{d}r \\
		+ \frac{2 q^2 n_0}{m \varepsilon_0} \int \limits_0^{\infty} \frac{e^{-r/\lambda}}{r} g(n_0, T_0, r) \Bigg( \left( 1 + \frac{r}{\lambda} + \frac{r^2}{3\lambda^2} \right) \\
		\times \left( -\frac{\sin(|\vec{k}|r)}{|\vec{k}|r} - \frac{3\cos(|\vec{k}|r)}{|\vec{k}|^2r^2} + \frac{3\sin(|\vec{k}|r)}{|\vec{k}|^3r^3} \right) + \frac{r^2}{6\lambda^2} \left( 1 - \frac{\sin(|\vec{k}|r)}{|\vec{k}|r} \right) \Bigg) \mathrm{d}r.
	\end{gathered}
\end{equation}
Here and below, the subscript ``$0$'' indicates evaluation at thermal equilibrium.
\par

As in the case of the OCP \cite{10.1063/5.0194352}, the variational principle predicts the existence of transverse waves.
The transverse dispersion law in Eq.~\eqref{eq:general_transverse_law} reduces to the OCP result in the limit of weak screening, $\lambda \to \infty$.
For finite screening, it agrees exactly with the QLCA results \cite{10.1063/1.873814, PhysRevLett.84.6030, 10.1063/1.4942169}, where the transverse modes of the YOCP have been extensively studied \cite{10.1063/1.873814, PhysRevLett.84.6030, 10.1063/1.4942169, 10.1063/1.5088141}.
We therefore focus our attention on the longitudinal mode.
\par

In contrast to the transverse case, the longitudinal dispersion depends not only on the pair distribution function but also on its derivatives.  
This mirrors the usual hydrodynamic results, where dispersion depends on derivatives of pressure, which itself is a derivative of energy.  
In our case, the internal energy includes both the microscopic kinetic contribution of an ideal gas and the interaction energy proportional to the pair distribution function, as given by Eq.~\eqref{eq:equilibrium_energy}.
The first term in Eq.~\eqref{eq:general_longitudinal_law}, which is proportional to $|\vec{k}|^2$, reflects this effect and generalizes the ideal gas contribution by incorporating correlation effects through derivatives of the pair distribution function.
Additional $|\vec{k}|$-dependent terms arise from extending hydrodynamics to nonlocal interactions and may also contribute at order $|\vec{k}|^2$ in a series expansion.
\par 

In the limit $\lambda \to \infty$, the longitudinal dispersion law reproduces the OCP result \cite{10.1063/5.0194352}.
For finite $\lambda$, the last integral in Eq.~\eqref{eq:general_longitudinal_law}, which depends on $g(n_0, T_0, r)$, matches the QLCA result \cite{10.1063/1.873814, PhysRevLett.84.6030, 10.1063/1.4942169}.
However, QLCA does not account for the thermal effects arising from both the kinetic energy and the temperature dependence of the pair distribution function.
\par

Thermodynamics of the YOCP is characterized by two dimensionless parameters \cite{PhysRevE.91.023108, Khrapak_2016}: the coupling parameter $\Gamma = q^2/(4\pi \varepsilon_0 a k_B T)$ and the screening parameter $\kappa = a/\lambda$.
It is therefore natural to express the longitudinal dispersion law in terms of these parameters.  
To do so, we use the fact that the pair distribution function, the equilibrium energy $E_0$, and the equilibrium pressure $p_0$ all depend on the dimensionless parameters $\Gamma$ and $\kappa$ as \cite{PhysRevE.91.023108, Khrapak_2016}
\begin{equation}
	\label{eq:thermodynamics_of_YOCP}
	\begin{gathered}
		g(n, T, r) = g \left( \Gamma, \kappa, \frac{r}{a} \right), \quad \frac{E_0}{Nk_B T} = \frac{3}{2} + u_{\mathrm{ex}}(\Gamma, \kappa), \quad \frac{p_0}{n k_B T} = 1 + p_{\mathrm{ex}}(\Gamma, \kappa).
	\end{gathered}
\end{equation}	
Here, $u_{\mathrm{ex}}$ and $p_{\mathrm{ex}}$ denote the dimensionless excess internal energy and excess pressure, respectively, which correspond to interaction-induced corrections to the ideal gas quantities.
\par 

Unlike in the case of the OCP \cite{Teller_1966}, the dimensionless excess energy and pressure in the YOCP are not simply proportional.
Nevertheless, they are related through the Helmholtz free energy \cite{PhysRevE.91.023108, Khrapak_2016}.  
Using thermodynamic identities \cite{landau2013statistical}, and the expressions for $E_0$ and $p_0$ in Eq.~\eqref{eq:thermodynamics_of_YOCP}, we can derive the adiabatic derivative of the temperature \cite{10.1063/5.0194352}.
The relevant partial derivatives are $\partial \Gamma/\partial T|_n = -\Gamma/T$, $\partial \Gamma/ \partial n|_T = \Gamma / (3n)$, $\partial \kappa/ \partial T|_n = 0$, and $\partial \kappa/\partial n|_T = -\kappa/(3n)$.
Using these, we obtain an expression 
\begin{equation}
	\label{eq:f1}
	\begin{gathered}
		\frac{n}{T} \frac{\partial T}{\partial n} \bigg|_{s} = \frac{2}{3} \frac{\left( 1 + p_{\mathrm{ex}}(\Gamma, \kappa) + \frac{1}{3}\left( \frac{\partial u_{\mathrm{ex}}}{\partial \kappa} \big|_{\Gamma} \kappa - \frac{\partial u_{\mathrm{ex}}}{\partial \Gamma} \big|_{\kappa} \Gamma \right) \right)}{\left( 1 - \frac{2\Gamma^2}{3} \frac{\partial (u_{\mathrm{ex}}/\Gamma)}{\partial \Gamma} \big|_{\kappa} \right)} = \frac{2}{3} f(\Gamma, \kappa),
		\end{gathered}
\end{equation}
consistent with the OCP result \cite{10.1063/5.0194352} when taking the limit of no $\kappa$ dependence and assuming $p_{\textrm{ex}} = u_{\textrm{ex}}/3$.
\par

We now define the plasma frequency, $\omega_p$, as $\omega_p^2 = q^2 n_0/(m \varepsilon_0)$, and introduce the dimensionless wave vector $\vec{q} = \vec{k}a$.
Motivated by scaling of the pair distribution function in Eq.~\eqref{eq:thermodynamics_of_YOCP}, we change the integration variable to $x = r/a$.
This allows us to rewrite the longitudinal dispersion relation from Eq.~\eqref{eq:general_longitudinal_law} as 
\begin{equation}
	\label{eq:longitudinal_law_YOCP}
	\begin{gathered}
		\left( \frac{\omega_L}{\omega_p} \right)^2 (|\vec{q}_0|) = \left(\frac{|\vec{q}_0|^2}{\Gamma_0} 
		- 2\Gamma_0 \frac{\partial j}{\partial \Gamma_0} \bigg|_{\kappa_0, |\vec{q}_0|} \right) \frac{F(\Gamma_0, \kappa_0)}{3} \\
		+ \frac{4}{9}f^2(\Gamma_0, \kappa_0) \frac{\partial}{\partial \Gamma_0} \bigg|_{\kappa_0, |\vec{q}_0|} \left( \Gamma_0^2 \frac{\partial j}{\partial \Gamma_0} \bigg|_{\kappa_0, |\vec{q}_0|} \right) \\
		- \frac{2}{3} f(\Gamma_0, \kappa_0) \Gamma_0 \frac{\partial \ell}{\partial \Gamma_0} \bigg|_{\kappa_0, |\vec{q}_0|} + b\left( \Gamma_0, \kappa_0, |\vec{q}_0| \right),
	\end{gathered}
\end{equation}
where the functions $f, F, j, \ell,$ and $b$ are defined by
\begin{equation}
	\label{eq:definition_F}
	\begin{gathered}
		F(\Gamma, \kappa) = f(\Gamma, \kappa)+\frac{2}{3}f^2(\Gamma, \kappa) + \frac{1}{3} \left( \frac{\partial f}{\partial \Gamma} \bigg|_{\kappa} \Gamma \left( 1 - 2f(\Gamma, \kappa) \right) -\frac{\partial f}{\partial \kappa} \bigg|_{\Gamma} \kappa \right),
	\end{gathered}
\end{equation}
\begin{equation}
	\label{eq:definition_j}
	\begin{gathered}
		j \left(\Gamma, \kappa, |\vec{q}| \right) = \frac{|\vec{q}|^2}{2} \int \limits_0^{\infty} xe^{-\kappa x} g(\Gamma, \kappa, x) \mathrm{d}x,
	\end{gathered}
\end{equation}
\begin{equation}
	\label{eq:definition_ell}
	\begin{gathered}
		\ell \left(\Gamma, \kappa, |\vec{q}| \right) = \int \limits_0^{\infty} \frac{e^{-\kappa x}}{x} \left( 1 + \kappa x \right) g(\Gamma, \kappa, x) \left( \frac{\sin \left(|\vec{q}|x \right)}{|\vec{q}|x} - \cos \left( |\vec{q}|x \right) \right) \mathrm{d}x,
	\end{gathered}
\end{equation}
\begin{equation}
	\label{eq:definition_b}
	\begin{gathered}
		b \left(\Gamma, \kappa, |\vec{q}| \right) = 2 \int \limits_0^{\infty} \frac{e^{-\kappa x}}{x} g(\Gamma, \kappa, x) \bigg\{ \left( 1 + \kappa x + \frac{\kappa^2 x^2}{3} \right) \bigg( - \frac{\sin \left(|\vec{q}|x \right)}{|\vec{q}|x} \\
		- 3\frac{\cos \left(|\vec{q}|x \right)}{|\vec{q}|^2x^2} + 3 \frac{\sin \left(|\vec{q}|x \right)}{|\vec{q}|^3x^3} \bigg) + \frac{\kappa^2 x^2}{6} \left(1 - \frac{\sin \left(|\vec{q}|x \right)}{|\vec{q}|x}\right) \bigg\} \mathrm{d}x.
	\end{gathered}
\end{equation}
The notation is chosen to maintain consistency with the OCP limit described in Ref.~\onlinecite{10.1063/5.0194352}.
\par 

\section{Thermodynamics and the step function approximation for the pair distribution function}
\label{section:thermodynamics_and_step_function_approximation}

The dispersion law derived in normalized variables can, in principle, be evaluated for any given pair distribution function. 
However, not only must the values of the pair distribution function be known accurately, but so must its derivatives with respect to $\Gamma$.  
As in the OCP case \cite{10.1063/5.0194352}, to the best of our knowledge, no such fits with reliable derivatives are currently available in the literature.
\par 

On the other hand, also as in the OCP case, the values of the dimensionless excess internal energy for the YOCP are well known from numerical simulations \cite{PhysRevE.56.4671, Caillol_2000, PhysRevE.90.053101}.  
In our calculations, we use the simple parameterization \cite{PhysRevE.91.023108, Khrapak_2016} 
\begin{equation}
	\label{eq:energy_fit}
	\begin{gathered}
		u_{\mathrm{ex}}(\Gamma, \kappa) = \frac{\kappa(\kappa + 1)\Gamma}{(\kappa+1)+(\kappa-1)e^{2\kappa}} + 3.2\left( \frac{\Gamma}{\Gamma_m(\kappa)} \right)^{2/5} - 0.1, 
	\end{gathered}
\end{equation}
which is accurate for $\kappa < 5$ and $10^{-3} < \Gamma/\Gamma_m(\kappa) < 1$. 
Here, $\Gamma_m(\kappa)$ is the coupling parameter at the fluid–solid phase transition (melting), and we adopt a simple parameterization \cite{Vaulina_2000, PhysRevE.66.016404}
\begin{equation}
	\label{eq:melting_fit}
	\begin{gathered}
		\Gamma_m(\kappa) = 172 \frac{e^{\alpha \kappa}}{\left(1 + \alpha \kappa + \frac{\alpha^2 \kappa^2}{2}\right)},
	\end{gathered}
\end{equation}
where $\alpha = (4\pi/3)^{1/3}$. We note that the screening dependence of the above expression for the melting line compares well with a parametrization of the effective coupling strength in Yukawa systems~\cite{10.1063/1.4900625}.
\par 

To compute the adiabatic derivative given in Eq.~\eqref{eq:f1}, which is relevant for the dispersion law, we also require an expression for the excess pressure.
However, no simple expression valid over the same range of $\Gamma$ and $\kappa$ appears to be available in the literature.
One possible approach is to perform particle-based numerical simulations and use particle positions and forces to compute the pressure using the virial equation \cite{Hansen_McDonald_2005}, followed by fitting to a simple form, as is done for the internal energy.  
Alternatively, one can calculate the equilibrium pair distribution function and then compute the pressure using the virial equation \cite{Hansen_McDonald_2005}.  
Although this approach has been shown to be accurate \cite{PhysRevE.90.053101}, it is not practical here, since we are unaware of any simple yet accurate parameterizations for the pair distribution function.
\par

Instead, in this paper we adopt an approach based on thermodynamic identities applied to the Helmholtz free energy \cite{PhysRevE.91.023108, Khrapak_2016, PhysRevE.56.4671, Caillol_2000, PhysRevE.90.053101}.  
The free energy is computed from the internal energy, for which simple fits are available.  
By differentiating it, we obtain a practical and accurate expression for the pressure.
However, directly integrating the fit in Eq.~\eqref{eq:energy_fit} leads to a divergence as $\Gamma \to 0^+$.  
To address this, we divide the analysis into two regimes \cite{PhysRevE.91.023108, Khrapak_2016, PhysRevE.56.4671, PhysRevE.90.053101}: weak coupling with $\Gamma < 1$ and strong coupling with $\Gamma > 1$.  
In the weakly coupled regime, the excess free energy is approximated using the second virial coefficient \cite{landau2013statistical}, which is accurate for $\kappa < 5$ \cite{Khrapak_2016}.  
For the strongly coupled regime, we use the fit in Eq.~\eqref{eq:energy_fit}.
\par 

The resulting expression for the dimensionless excess pressure is given by
\begin{equation}
	\label{eq:pressure_fit}
	\begin{gathered}
		p_{\mathrm{ex}}(\Gamma, \kappa) = \frac{\kappa}{2} \int \limits_0^{\infty} e^{-\exp(-\kappa x)/x} e^{-\kappa x} x^2 \mathrm{d}x + \frac{u_{\mathrm{ex}}(\Gamma, \kappa)}{3} \\
		+ \frac{\kappa(\Gamma-1)}{3} \frac{\left(e^{2 \kappa} \left(1 - \kappa^2 + 2\kappa^3 \right) - (\kappa + 1)^2 \right)}{\left( (\kappa+1)+(\kappa-1)e^{2\kappa} \right)^2} \\
		+ \frac{16}{15} \frac{\left(\Gamma^{2/5}-1 \right)}{\Gamma_m^{2/5}(\kappa)} \frac{\alpha^3 \kappa^3}{\left(2 + 2\alpha \kappa + \alpha^2 \kappa^2 \right)}.
	\end{gathered}
\end{equation}
The first term corresponds to the second virial coefficient, while the remaining terms tend to the known strongly coupled limit as $\Gamma \gg 1$ \cite{PhysRevE.91.023108}.  
This analytical formula also agrees well with results for the pressure derived from the equilibrium pair distribution function, as discussed in Ref.~\onlinecite{PhysRevE.90.053101}.  
By construction, the expression is valid only for $\Gamma > 1$.
\par

The longitudinal dispersion law in Eq.~\eqref{eq:longitudinal_law_YOCP} also requires knowledge of the pair distribution function, as seen in Eqs.~\eqref{eq:definition_j}, \eqref{eq:definition_ell}, and \eqref{eq:definition_b}.  
Since no precise fits are available, we use a simple step function approximation that has proven effective in the OCP case \cite{10.1063/5.0194352} and has also been used in QLCA applications to the YOCP \cite{10.1063/1.4942169, 10.1063/1.5088141}.  
More refined approaches, such as a two-step potential, are possible, but previous studies of QLCA have shown that they do not yield substantial improvements \cite{FAIRUSHIN2020103359}.
\par

In the step function approximation, we assume $g(n, T, r) = 0$ for $r < r_c(n,T)$ and $g(n, T, r) = 1$ otherwise. 
This form is consistent with theoretical limits as $r \to 0^+$ and $r \to \infty$ \cite{Teller_1966, Hansen_McDonald_2005}.  
Here, $r_c(n,T)$ is the cutoff radius at which the pair distribution function jumps discontinuously and is the only parameter required to define the approximation.
For the OCP, $r_c(n,T)$ was fixed by requiring that the internal energy computed under this approximation reproduces the known values from accurate thermodynamic calculations \cite{10.1063/5.0194352}.  
For the YOCP, because the excess pressure is not simply proportional to the excess internal energy, one could instead enforce agreement with pressure data, leading to a different $r_c(n,T)$ dependence \cite{10.1063/1.4942169}. 
However, QLCA studies have shown that whether energy or pressure is used to determine $r_c(n,T)$, the resulting dispersion curves are very similar \cite{10.1063/1.4942169}.  
Given this, and considering that internal energy is more commonly computed in numerical simulations and has simple fits available, we choose to determine $r_c(n,T)$ by matching the internal energy.
\par

The general expression for the excess internal energy from Eq.~\eqref{eq:equilibrium_energy} can be rewritten in dimensionless form and evaluated for the step function approximation \cite{10.1063/1.4942169}, leading to 
\begin{equation}
	\label{eq:energy_expression}
	\begin{gathered}
		u_{\mathrm{ex}}(\Gamma, \kappa) = \frac{3\Gamma}{2\kappa^2} e^{-\kappa x_c(\Gamma, \kappa)} \left( 1 + \kappa x_c(\Gamma, \kappa) \right),
	\end{gathered}
\end{equation}
where we define the dimensionless cutoff radius $x_c = r_c/a$, which depends on $\Gamma$ and $\kappa$.  
Matching this equation to Eq.~\eqref{eq:energy_fit} allows us to determine $x_c(\Gamma, \kappa)$ over the range of $\Gamma$ and $\kappa$ where the energy fit is valid.
\par

This approximation significantly simplifies the integrals appearing in the longitudinal dispersion relation, specifically those in Eqs.~\eqref{eq:definition_j}, \eqref{eq:definition_ell}, and \eqref{eq:definition_b}, which are now rewritten as 
\begin{equation}
	\label{eq:definition_j_simplified}
	\begin{gathered}
		j \left(\Gamma, \kappa, |\vec{q}| \right) = \frac{|\vec{q}|^2}{2\kappa^2} e^{-\kappa x_c} \left( 1 + \kappa x_c \right),
	\end{gathered}
\end{equation}
\begin{equation}
	\label{eq:definition_ell_simplified}
	\begin{gathered}
		\ell \left(\Gamma, \kappa, |\vec{q}| \right) = e^{-\kappa x_c} \left( \frac{\sin \left( |\vec{q}|x_c\right)}{|\vec{q}|x_c} + \frac{\kappa}{\left(\kappa^2 + |\vec{q}|^2\right)} \left( |\vec{q}| \sin \left( |\vec{q}|x_c \right) - \kappa \cos \left( |\vec{q}|x_c \right)  \right) \right),
	\end{gathered}
\end{equation}
\begin{equation}
	\label{eq:definition_b_simplified}
	\begin{gathered}
		b \left(\Gamma, \kappa, |\vec{q}| \right) = e^{-\kappa x_c} \bigg( \left( 1 + \kappa x_c \right) \left( \frac{1}{3} - 2\frac{\cos \left( |\vec{q}|x_c \right) }{|\vec{q}|^2 x_c^2} + 2 \frac{\sin \left( |\vec{q}|x_c \right) }{|\vec{q}|^3 x_c^3} \right)  \\
	- \frac{\kappa^2}{\left(\kappa^2 + |\vec{q}|^2\right)} \left( \cos \left( |\vec{q}|x_c \right) + \frac{\kappa}{|\vec{q}|} \sin \left( |\vec{q}|x_c \right)  \right) \bigg).
	\end{gathered}
\end{equation}
In these expressions, the dependence of $x_c$ on $\Gamma$ and $\kappa$ is implied but not written out explicitly.
\par

\section{Results and comparison with molecular dynamics and other theoretical approaches}
\label{section:results_and_comparison}

\begin{figure}
    \includegraphics{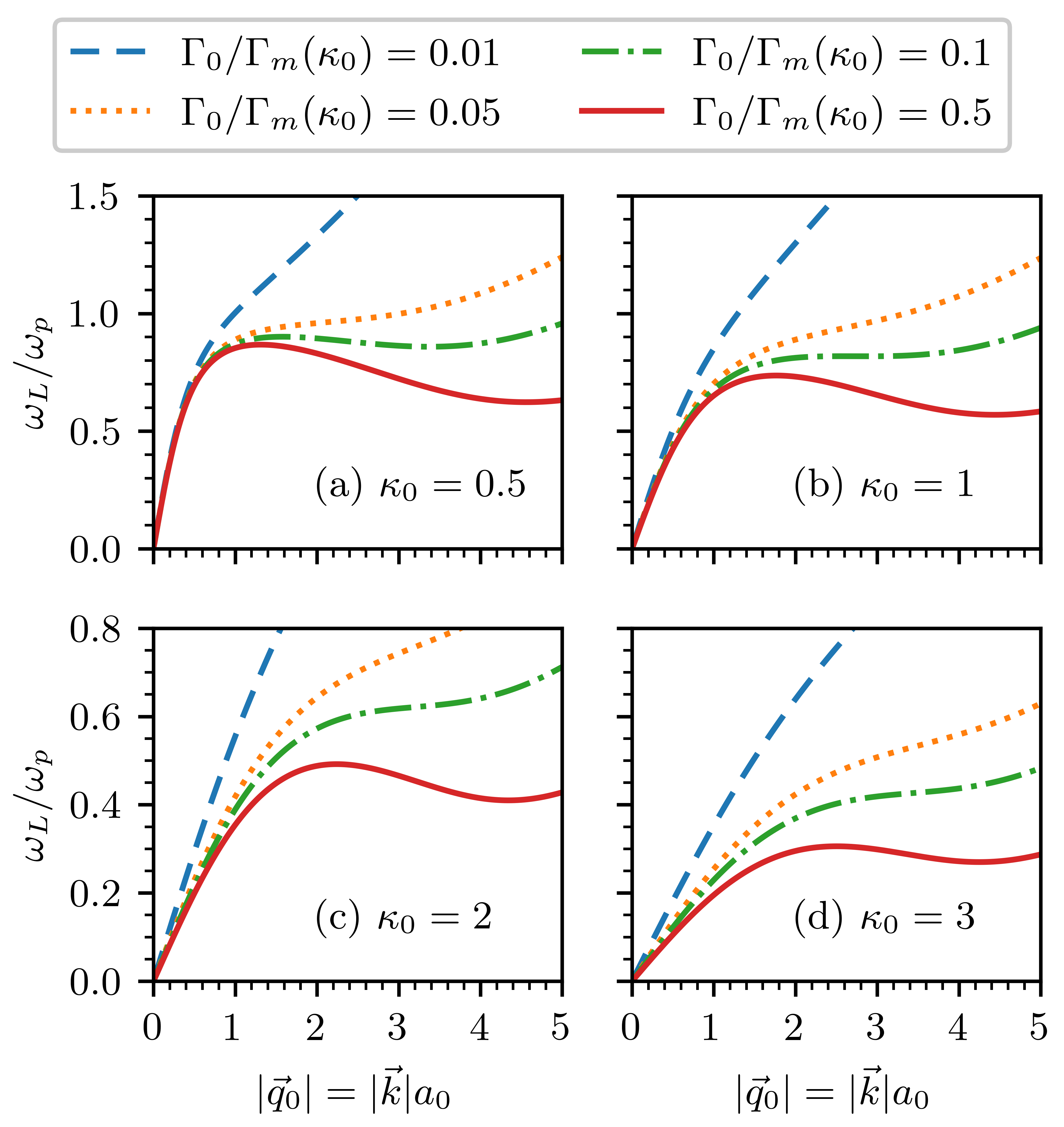}
	\caption{
		Longitudinal dispersion law in normalized variables for different values of $\Gamma_0$ and $\kappa_0$, as predicted by the variational principle. 
        The coupling parameter $\Gamma_0$ is scaled by the melting value $\Gamma_m(\kappa_0)$, estimated using Eq.~\eqref{eq:melting_fit}. 
		The dispersion law in Eq.~\eqref{eq:longitudinal_law_YOCP} is evaluated using the step function approximation for the pair distribution function.
        Eqs.~\eqref{eq:f1} and \eqref{eq:definition_F}, together with the simplified Eqs.~\eqref{eq:energy_expression}, \eqref{eq:definition_j_simplified}, \eqref{eq:definition_ell_simplified}, and \eqref{eq:definition_b_simplified} are used.
		}
	\label{fig:comparison_of_longitudinal}
\end{figure}

Before comparing the longitudinal dispersion relation predicted by our hydrodynamic theory with other theoretical results and MD simulations, we first discuss the general features of our predictions. 
Fig.~\ref{fig:comparison_of_longitudinal} shows the longitudinal dispersion law given by Eq.~\eqref{eq:longitudinal_law_YOCP}, evaluated using the step function approximation for the pair distribution function.
In this approximation, Eqs.~\eqref{eq:definition_j_simplified}, \eqref{eq:definition_ell_simplified}, and \eqref{eq:definition_b_simplified} are used, where the dimensionless cutoff radius is determined from Eq.~\eqref{eq:energy_expression}.
The functions $f$ and $F$, defined in Eqs.~\eqref{eq:f1} and \eqref{eq:definition_F}, are computed using the energy and pressure fits given by Eqs.~\eqref{eq:energy_fit}, \eqref{eq:melting_fit}, and \eqref{eq:pressure_fit}.
\par 

A general trend observed is that as $\kappa_0$ increases, the excitation frequencies decrease.
Nevertheless, the qualitative behavior with respect to the coupling parameter $\Gamma_0$ is similar across different values of $\kappa_0$. 
In particular, when $\Gamma_0$ is much smaller than the melting value, the dispersion relation increases monotonically with the magnitude of the wave vector.
However, as $\Gamma_0$ increases and approaches the melting threshold, the dispersion law develops both a local maximum and a local minimum in the range $0 \leq |\vec{q}_0| \leq 5$, with their precise locations depending on $\kappa_0$.
The position of the maximum shifts to higher $|\vec{q}_0|$ as $\kappa_0$ increases, while the position of the minimum shows only a weak dependence on $\kappa_0$.
The overall behavior as a function of $\Gamma_0$, and the emergence of a local minimum, qualitatively resemble the results for the OCP \cite{10.1063/5.0194352}, with the key difference that in the limit $|\vec{q}_0| \to 0^+$, we now have $\omega_L \to 0^+$, rather than $\omega_L \to \omega_p$.
\par 

\par 

\subsection{Comparison of the longitudinal speed of sound}

\begin{figure}
    \includegraphics{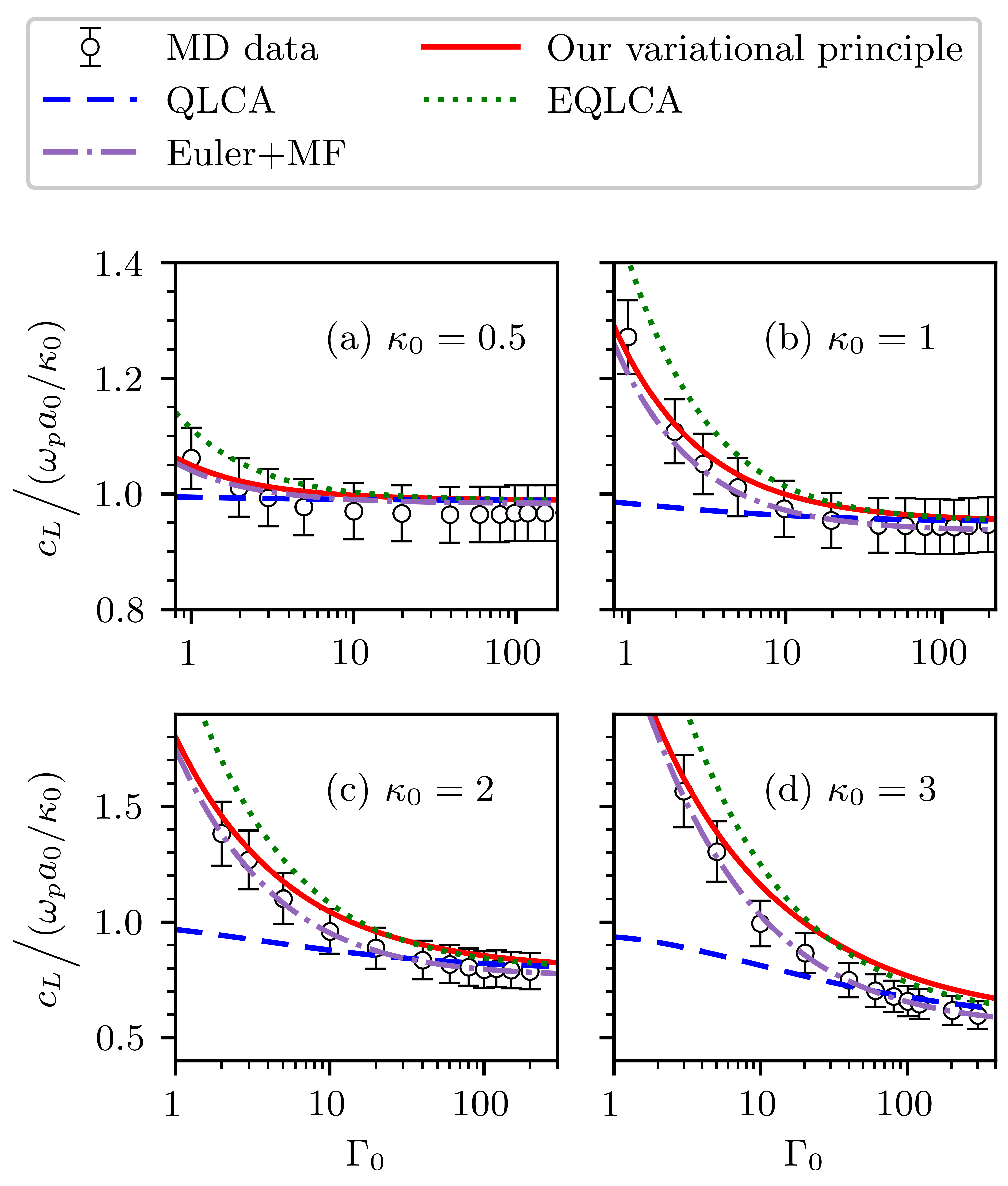}
	\caption{
        The dimensionless longitudinal speed of sound, as defined in Eq.~\eqref{eq:speed_of_sound_def}, for various values of $\Gamma_0$ and $\kappa_0$. 
        Circles with error bars denote MD data from Ref.~\onlinecite{PhysRevE.100.063206}, and the error bars reflect an uncertainty of $5\%$ for (a) and (b), and $10\%$ for (c) and (d).
        The solid red line corresponds to our variational approach using Eq.~\eqref{eq:longitudinal_law_YOCP} with the step function approximation.
        Eqs.~\eqref{eq:f1} and \eqref{eq:definition_F}, together with the simplified Eqs.~\eqref{eq:energy_expression}, \eqref{eq:definition_j_simplified}, \eqref{eq:definition_ell_simplified}, and \eqref{eq:definition_b_simplified} are used.
        The dashed blue line shows the QLCA result, while the dotted green line shows the extended QLCA (EQLCA), given by Eq.~\eqref{eq:extended_QLCA_dispersion}, both evaluated using the same step function approximation for the pair distribution function.
        The dot-dashed purple line corresponds to the Euler hydrodynamic theory with a mean field term, given by Eq.~\eqref{eq:HMF_dispersion}.
	}
	\label{fig:speed_of_sound}
\end{figure}

As a first test of the accuracy of our theory, we consider the longitudinal speed of sound in the long-wavelength limit. 
High-precision data for a wide range of $\Gamma_0$ and $\kappa_0$ are available from MD simulations \cite{PhysRevE.100.063206, PhysRevResearch.2.033287}. 
We define the longitudinal speed of sound, $c_L$, in dimensionless units consistent with those used in the literature as 

\begin{equation}
	\label{eq:speed_of_sound_def}
	\begin{gathered}
	\frac{c_L^2}{(\omega_p a_0/\kappa_0)^2} = \kappa_0^2 \lim \limits_{|\vec{q}_0| \to 0^+} \frac{(\omega_L/\omega_p)^2(|\vec{q}_0|)}{|\vec{q}_0|^2}
	= \left(\frac{1}{\Gamma_0} 
	+ \frac{\Gamma_0}{2} e^{-\kappa_0 x_c}\frac{\partial x_c^2}{\partial \Gamma_0} \bigg|_{\kappa_0} \right) \frac{\kappa_0^2 F(\Gamma_0, \kappa_0)}{3}
    \\
	- \frac{\kappa_0^2}{9} f^2(\Gamma_0, \kappa_0) \frac{\partial}{\partial \Gamma_0} \bigg|_{\kappa_0} \left( \Gamma_0^2 e^{-\kappa_0 x_c}\frac{\partial x_c^2}{\partial \Gamma_0} \bigg|_{\kappa_0} \right)
	+ \frac{\kappa_0^2}{9} f(\Gamma_0, \kappa_0)\Gamma_0 e^{-\kappa_0 x_c} (1+ \kappa_0 x_c) \frac{\partial x_c^2}{\partial \Gamma_0} \bigg|_{\kappa_0}
	\\
    + e^{-\kappa_0 x_c} \left( 1 + \kappa_0 x_c + \frac{13}{30} \kappa_0^2 x_c^2 + \frac{1}{10} \kappa_0^3 x_c^3 \right),
	\end{gathered}
\end{equation}
which for each value of $\Gamma_0$ and $\kappa_0$ was computed directly using the dispersion relation in Eq.~\eqref{eq:longitudinal_law_YOCP}.
Fig.~\ref{fig:speed_of_sound} presents the computed speed of sound along with MD results \cite{PhysRevE.100.063206}.
While simulation data for smaller $\kappa_0$ values exist even for smaller coupling with $\Gamma_0 < 1$, we restrict our analysis to $\Gamma_0 \geq 1$, as this is the valid range for the used equation of state given by Eqs.~\eqref{eq:energy_fit}, \eqref{eq:melting_fit}, and \eqref{eq:pressure_fit}.
\par

\par 

Our variational approach captures the qualitative dependence of the longitudinal speed of sound on $\Gamma_0$ for all $\kappa_0$ values.
Quantitative agreement is particularly good for $\kappa_0 \leq 1$.
For $\kappa_0 = 0.5$, the error at most reaches $3\%$, showing excellent agreement.
A possible explanation for the disagreement of all theories with the simulations at $\kappa_0=0.5$ and large $\Gamma_0$ could be 
related to difficulties in extracting the sound speed from finite wave number MD data.
In particular, for small $\kappa_0$, the linear part of the dispersion relation is shortest, making the results sensitive to the smallest wave vector used, which in the data of Ref.~\onlinecite{PhysRevE.100.063206} is $|\vec{q}_0| = 0.1289$. This may explain why the theoretical curves overestimate the numerical data.
For $\kappa_0 = 1.0$, the agreement of our hydrodynamic model and the simulations is excellent over the entire range of considered $\Gamma_0$. 
The peak error of $3\%$ is achieved at moderate coupling $\Gamma_0 = 20 = 0.09\Gamma_m(\kappa_0)$.
At stronger screening, where $\kappa_0 \geq 2$, deviations become more noticeable.
For $\kappa_0 = 2$, the maximum error reaches $9\%$ at $\Gamma_0 = 10 = 0.02\Gamma_m(\kappa_0)$, and for $\kappa_0 = 3$, it reaches $17\%$ at $\Gamma_0 = 40 = 0.03\Gamma_m(\kappa_0)$.
While such discrepancies might be attributed to the equation of state or the step function approximation for the pair distribution function, our own MD simulations, which remove both approximations, together with the dispersion law given by Eq.~\eqref{eq:longitudinal_law_YOCP}, indicate that their effect is around $1\%$, suggesting another source for the disagreement.
We have also verified, using the finite wave vector dispersion relations, that the smallest wave vector, $|\vec{q}_0|=0.1289$, used in the MD data has no significant impact at such high screening.
\par 

We also compare our results with those from the QLCA, a well-established analytical method for obtaining dispersion relations in strongly coupled systems \cite{PhysRevA.41.5516, 10.1063/1.873814, Donkó_2008}.
This method has been extensively studied in the literature for both the OCP and YOCP, and provides excellent results in the strong coupling regime, $\Gamma_0 \gg 1$ \cite{Donkó_2008, PhysRevLett.84.6030, 10.1063/1.4942169, PhysRevE.100.063206}.
Within this framework, the longitudinal dispersion relation corresponds to the last term in Eq.~\eqref{eq:longitudinal_law_YOCP}, specifically given in Eq.~\eqref{eq:definition_b}.
For consistency, we evaluate it using the step function approximation for the pair distribution function, as discussed earlier.
In the context of QLCA, this approximation has been shown to be accurate in the long-wavelength regime, especially when compared to more precise representations of the pair distribution function~\cite{10.1063/1.4942169}.
The resulting speed of sound is shown in Fig.~\ref{fig:speed_of_sound}.
\par 

At strong coupling, with $\Gamma_0 \gg 1$, the QLCA agrees very well with simulation data across all values of $\kappa_0$, and it outperforms our theory for $\kappa_0 = 2$ and $\kappa_0 = 3$.
While our theory converges to the QLCA result as $\Gamma_0$ increases, this convergence is slower under strong screening and insufficient to match the numerical data.
Nevertheless, the strength of our approach lies in its improved performance at moderate and weak coupling, due to its inclusion of thermal effects, and in its applicability to nonlinear problems.
\par 

To incorporate thermal effects into the QLCA dispersion law, particularly important at low values of the coupling parameter $\Gamma_0$, one can add a kinetic term either phenomenologically \cite{PhysRevE.79.046412, PhysRevE.87.043102, 10.1063/1.4965903} or by deriving it from kinetic theory \cite{https://doi.org/10.1002/ctpp.202400018}.
With this modification, the dispersion relation in the extended QLCA (EQLCA) framework becomes 
\begin{equation}
	\label{eq:extended_QLCA_dispersion}
	\begin{gathered}
		\left( \frac{\omega_{L, \textrm{EQCLA}}}{\omega_p} \right)^2 (|\vec{q}_0|) = \frac{|\vec{q}_0|^2}{\Gamma_0} + b\left( \Gamma_0, \kappa_0, |\vec{q}_0| \right),
	\end{gathered}
\end{equation}
where $b$ is defined in Eq.~\eqref{eq:definition_b}.
For consistency, it is computed using the step function approximation for the pair distribution function.
\par 

Fig.~\ref{fig:speed_of_sound} shows the resulting speed of sound.
As expected, the EQLCA converges to the standard QLCA result as $\Gamma_0$ increases, while providing improved behavior at lower coupling.
For $\kappa_0 = 0.5$, the agreement with MD data is excellent.
At larger values of $\kappa_0$, the discrepancy increases, similar to what is observed with our variational approach.
However, in the regime of strong screening and weak coupling, our variational theory offers a better description of the speed of sound.
In the limit $\Gamma_0 \gg 1$, the EQLCA and our approach yield closely matching results.
\par

For the final comparison, we consider an approach based on the standard Euler equations of hydrodynamics augmented by a mean-field term (Euler+MF), which accounts for the screened interaction potential.
This framework can be formulated using a fluid description for single-component~\cite{10.1063/1.2759881} or multi-component plasmas~\cite{RAO1990543, PhysRevE.91.033110, Khrapak_2016}.
The resulting longitudinal dispersion law corresponds to the combination of the mean-field term and the adiabatic speed of sound.
Expressed in terms of the equation of state for the YOCP without background, as defined by Eqs.~\eqref{eq:thermodynamics_of_YOCP}, \eqref{eq:energy_fit}, \eqref{eq:melting_fit}, and \eqref{eq:pressure_fit}, the dispersion relation becomes
\begin{equation}
    \label{eq:HMF_dispersion_k_space}
    \omega_{L, \textrm{Euler+MF}}^2 = \left( \frac{1}{m} \frac{\partial p_0}{\partial n_0} \bigg|_{s_0} + \frac{\omega_p^2}{|\vec{k}|^2 + (1/\lambda)^2} - \omega_p^2 \lambda^2 \right) |\vec{k}|^2.
\end{equation}
\par 

In the long-wavelength limit, where $|\vec{k}| \to 0^+$, this expression reduces to $\omega_{L, \textrm{Euler+MF}}^2 \approx (1/m)$ $(\partial p_0/\partial n_0)|_{s_0} |\vec{k}|^2$, which corresponds to the classical hydrodynamic result without explicit interaction terms \cite{landau1987fluid}.
In this limit, interactions enter only through the equation of state.
The general form, as given in Eq.~\eqref{eq:HMF_dispersion_k_space}, in normalized variables is given by
\begin{equation}
	\label{eq:HMF_dispersion}
	\begin{gathered}
		\left( \frac{\omega_{L, \textrm{Euler+MF}}}{\omega_p} \right)^2 (|\vec{q}_0|) = \frac{|\vec{q}_0|^2}{3 \Gamma_0} \Bigg( \left( 1 + p_{\mathrm{ex}}(\Gamma_0, \kappa_0) \right) \left( 1 + \frac{2}{3} f(\Gamma_0, \kappa_0) \right) \\
        + \frac{1}{3} \left( \frac{\partial p_{\mathrm{ex}}}{\partial \Gamma_0} \bigg|_{\kappa_0} \Gamma_0 \left( 1 - 2f(\Gamma_0, \kappa_0) \right) -\frac{\partial p_{\mathrm{ex}}}{\partial \kappa_0} \bigg|_{\Gamma_0} \kappa_0 \right) \Bigg)
        + \frac{|\vec{q}_0|^2}{|\vec{q}_0|^2 + \kappa_0^2} - \frac{|\vec{q}_0|^2}{\kappa_0^2}. 
	\end{gathered}
\end{equation}
\par 

Fig.~\ref{fig:speed_of_sound} includes the results for the speed of sound computed using this approach.
For screening parameters $\kappa_0 \leq 1$, the Euler+MF model agrees very well with the MD data, with deviations comparable in magnitude to those of our variational approach.
However, for stronger screening with $\kappa_0 = 2$ and $\kappa_0 = 3$, this method continues to show excellent agreement with the simulations, unlike our variational method, which begins to deviate.
\par 

The fact that the Euler+MF model reproduces simulation data well at small wave vectors is well established \cite{PhysRevE.100.063206, PhysRevE.83.015401}.
Nevertheless, our variational approach provides a more accurate description on length scales comparable to the average interparticle spacing.
We demonstrate this in the following subsection through a comparison with MD simulations at normalized wave vector magnitudes on the order of unity.
In addition, our approach has the advantage of being applicable to both the YOCP and the OCP~\cite{10.1063/5.0194352}, whereas the Euler+MF model is known to fail for the OCP, even in the long-wavelength limit where it would be expected to perform well~\cite{PhysRevE.83.015401, 10.1063/5.0229805}.
\par 

\subsection{Comparison of the longitudinal dispersion law at finite wavelength}

One of the advantages of the hydrodynamic equations derived from our variational approach is their ability to describe plasma behavior down to short length scales, comparable to the average interparticle spacing.
To verify the accuracy at such length scales, we performed our own MD simulations using the LAMMPS code~\cite{LAMMPS}. 
They were conducted with $N=10^4$ particles and a time step of $\omega_p \Delta t =0.01$. 
With the trajectories $\{\vec r_a(t)\}$, where $a\in\{1,\dots,N\}$, we computed the dynamic structure factor,~\cite{kaehlert2019, PhysRevResearch.2.033287}
\begin{equation}
S(|\vec q_0|,\omega)=\frac{1}{2\pi}\int_{-\infty}^\infty F(|\vec q_0|,t)e^{i\omega t}\text{d}t,
\end{equation}
where $F(|\vec q_0|,t)=\langle n(\vec q_0,t)n(-\vec q_0,0) \rangle /N$ is the density auto-correlation function and $n(\vec q_0,t)=\sum_a e^{-i\vec q_0\cdot \vec r_a(t)}$ the microscopic density. 
From the dynamic structure factor, we obtained the longitudinal current fluctuation spectrum, $L(|\vec{q}_0|, \omega) = \omega^2 S(|\vec{q}_0|, \omega)$.
To visualize the peaks that indicate the longitudinal dispersion relation, we normalized the spectrum at each $|\vec{q}_0|$ by its maximum value.
\par 

\begin{figure}
    \includegraphics{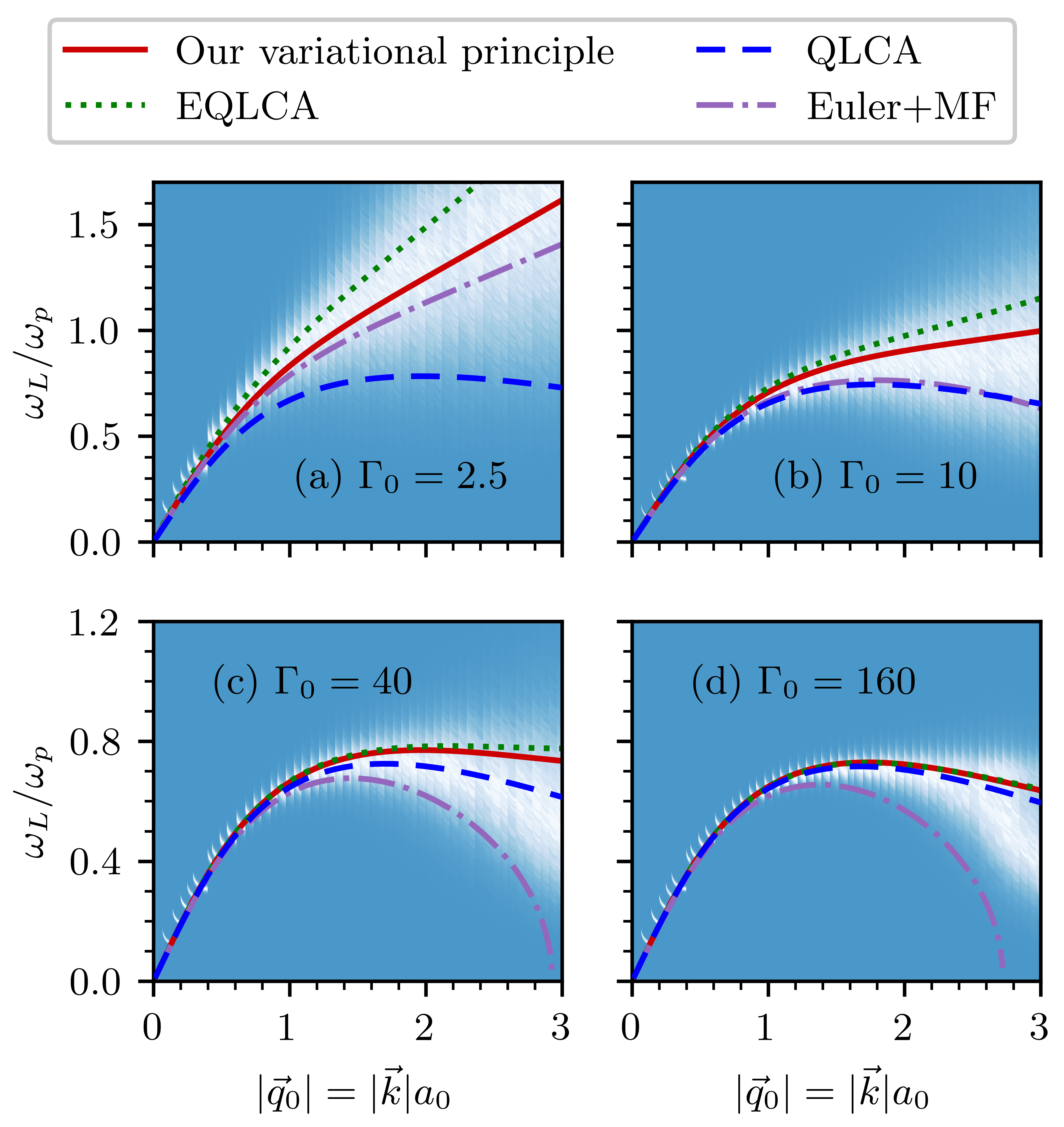}
	\caption{
        The longitudinal dispersion law at finite wavelengths in normalized variables for $\kappa_0 = 1$ and $\Gamma_0 = 2.5, 10, 40, 160$, corresponding approximately to $\Gamma_0/\Gamma_m(\kappa_0) = 0.01, 0.05, 0.18, 0.73$, where $\Gamma_m(\kappa_0) = 220$ is the coupling parameter at melting, as given by Eq.~\eqref{eq:melting_fit}.
        The colored background represents results from our MD simulations of the longitudinal current fluctuation spectrum, normalized at each $|\vec{q}_0|$ value, and the white regions indicate peak locations.
        The solid red line corresponds to our variational approach using Eq.~\eqref{eq:longitudinal_law_YOCP} with the step function approximation.
        Eqs.~\eqref{eq:f1} and \eqref{eq:definition_F}, together with the simplified Eqs.~\eqref{eq:energy_expression}, \eqref{eq:definition_j_simplified}, \eqref{eq:definition_ell_simplified}, and \eqref{eq:definition_b_simplified} are used.
        The dashed blue line shows the QLCA result, while the dotted green line shows the extended QLCA (EQLCA), given by Eq.~\eqref{eq:extended_QLCA_dispersion}, both evaluated using the same step function approximation for the pair distribution function.
        The dot-dashed purple line corresponds to the Euler hydrodynamic theory with a mean field term, given by Eq.~\eqref{eq:HMF_dispersion}.
	}
	\label{fig:comparison_of_finite_wavelength_kappa_1}
\end{figure}

\begin{figure}
    \includegraphics{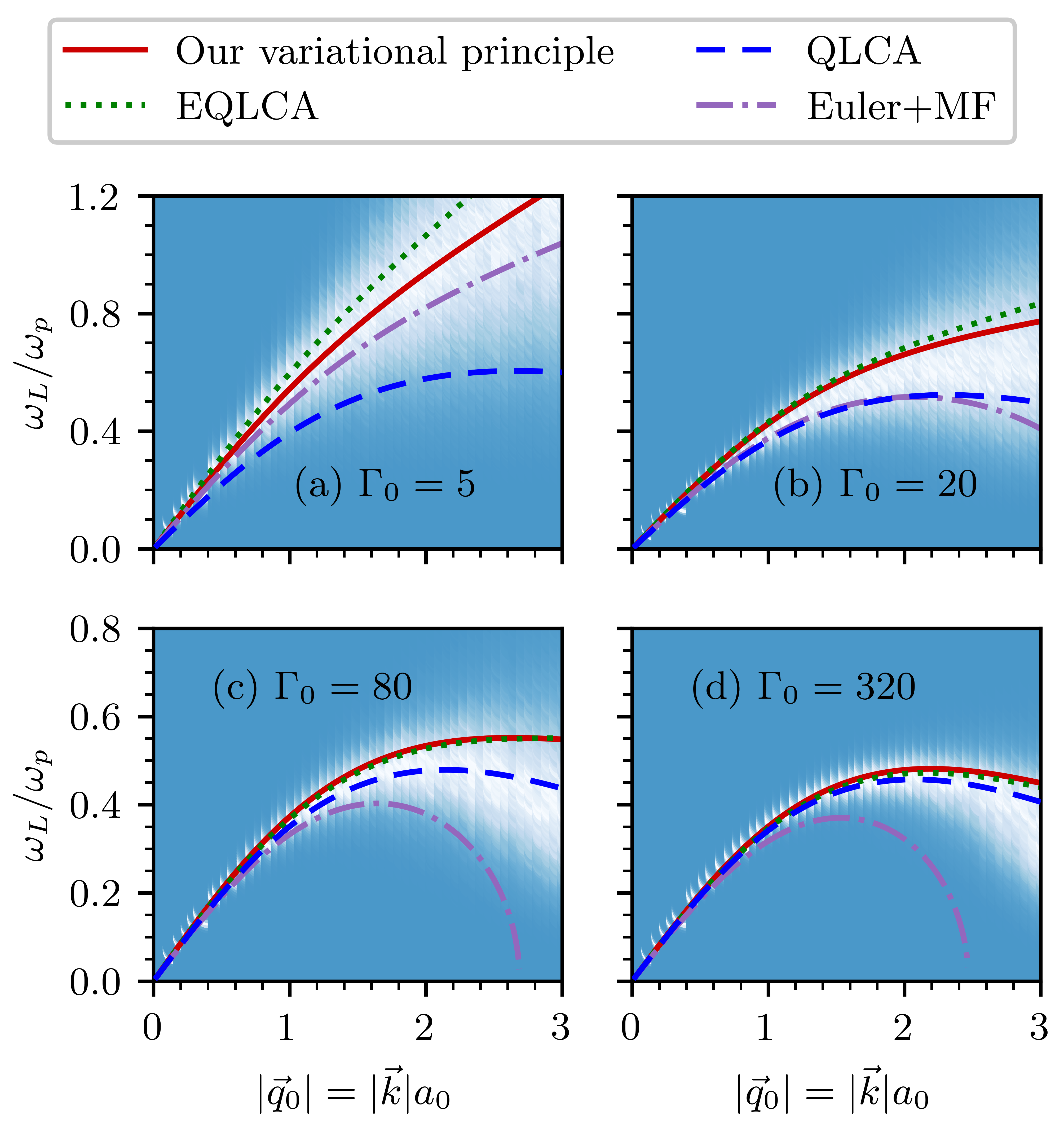}
	\caption{
        The longitudinal dispersion law at finite wavelengths in normalized variables for $\kappa_0 = 2$ and $\Gamma_0 = 5, 20, 80, 320$, corresponding approximately to $\Gamma_0/\Gamma_m(\kappa_0) = 0.01, 0.04, 0.17, 0.70$, where $\Gamma_m(\kappa_0) = 460$ is the coupling parameter at melting, as given by Eq.~\eqref{eq:melting_fit}.
        The colored background represents results from our MD simulations of the longitudinal current fluctuation spectrum, normalized at each $|\vec{q}_0|$ value, and the white regions indicate peak locations.
        The solid red line corresponds to our variational approach using Eq.~\eqref{eq:longitudinal_law_YOCP} with the step function approximation.
        Eqs.~\eqref{eq:f1} and \eqref{eq:definition_F}, together with the simplified Eqs.~\eqref{eq:energy_expression}, \eqref{eq:definition_j_simplified}, \eqref{eq:definition_ell_simplified}, and \eqref{eq:definition_b_simplified} are used.
        The dashed blue line shows the QLCA result, while the dotted green line shows the extended QLCA (EQLCA), given by Eq.~\eqref{eq:extended_QLCA_dispersion}, both evaluated using the same step function approximation for the pair distribution function.
        The dot-dashed purple line corresponds to the Euler hydrodynamic theory with a mean field term, given by Eq.~\eqref{eq:HMF_dispersion}.
	}
	\label{fig:comparison_of_finite_wavelength_kappa_2}
\end{figure}

Figs.~\ref{fig:comparison_of_finite_wavelength_kappa_1} and \ref{fig:comparison_of_finite_wavelength_kappa_2} show the comparison for $\kappa_0 = 1$ and $\kappa_0 = 2$, respectively, between our variational approach, our MD data for the longitudinal current fluctuation spectrum, and other theoretical approaches: QLCA, EQCLA, as given in Eq.~\eqref{eq:extended_QLCA_dispersion}, and Euler+MF, as given in Eq.~\eqref{eq:HMF_dispersion}.
The comparison is restricted to $|\vec{q}_0| \leq 3$, as the spectral peaks become significantly broader at larger wave vectors, and dissipative processes begin to play an important role. 
However, these processes are not currently incorporated into our variational framework.

\par 

From Fig.~\ref{fig:comparison_of_finite_wavelength_kappa_1}, corresponding to $\kappa_0 = 1$, we observe that over a wide range of coupling parameters, from $\Gamma_0 = 2.5 = 0.01\Gamma_m(\kappa_0)$ up to $\Gamma_0 = 160 = 0.73\Gamma_m(\kappa_0)$, our variational approach accurately reproduces the peak structure of the longitudinal current fluctuation spectrum from simulations. 
Remarkably, this agreement holds even for wavelengths on the order of the average interparticle spacing.
At weak coupling, where $\Gamma_0 = 2.5 = 0.01\Gamma_m(\kappa_0)$ and $\Gamma_0 = 10 = 0.05\Gamma_m(\kappa_0)$, our approach provides the best match to the numerical data compared to other theories. 
At moderate coupling, corresponding to $\Gamma_0 = 40 = 0.18\Gamma_m(\kappa_0)$, QLCA yields a more accurate prediction, while our model slightly overestimates the dispersion relation at large wave vectors, approaching the EQLCA result. 
At strong coupling, corresponding to $\Gamma_0 = 160 = 0.73\Gamma_m(\kappa_0)$, our theory, QLCA, and EQLCA all converge.
Also, the Euler+MF model underestimates the dispersion relation at low coupling, while at large values of the coupling parameter, it predicts regions of instability at high $|\vec{q}_0|$, where $\omega_L^2 < 0$.
Overall, across the full range of coupling strengths, our variational model provides the best match to MD results.+
\par 

In the case of $\kappa_0 = 2$, as shown in Fig.~\ref{fig:comparison_of_finite_wavelength_kappa_2}, similar conclusions hold. 
Across a wide coupling range, from $\Gamma_0 = 5 = 0.01\Gamma_m(\kappa_0)$ to $\Gamma_0 = 320 = 0.70\Gamma_m(\kappa_0)$, our model shows reasonable agreement with the MD data, even at wavelengths comparable to the average interparticle spacing.
At weak coupling, corresponding to $\Gamma_0 = 5 = 0.01\Gamma_m(\kappa_0)$, our model again outperforms other approaches.
Increasing the coupling to $\Gamma_0 = 20 = 0.04\Gamma_m(\kappa_0)$, the dispersion is slightly overestimated but remains better than for other approaches.
At moderate coupling, corresponding to $\Gamma_0 = 80 = 0.17\Gamma_m(\kappa_0)$, QLCA aligns best with the data, while our theory continues to overestimate the dispersion relation. Finally, at strong coupling with $\Gamma_0 = 320 = 0.70\Gamma_m(\kappa_0)$, our approach again converges to both QLCA and EQLCA.
As before, the Euler+MF model underestimates the dispersion relation at low coupling and at high coupling exhibits regions of instability at large $|\vec{q}_0|$.
\par 

Notably, at strong screening, our theory tends to align with EQLCA across most of the coupling range, except at very low $\Gamma_0$.
Nevertheless, our approach maintains an advantage over EQLCA in that it is applicable to nonlinear problems.
As in the case of $\kappa_0 = 1$, if we consider the full coupling range rather than isolated values, our variational theory yields the most consistent agreement with MD simulations among all theoretical models considered.
\par 

\section{Conclusions}
\label{section:conclusions}

We have applied a previously developed variational principle~\cite{10.1063/5.0194352} to the YOCP and derived hydrodynamic equations that incorporate correlation and strong coupling effects. 
We derived explicit equations of motion in both reference coordinates, corresponding to a reference state such as the initial state or a state of thermal equilibrium, and laboratory coordinates, together with the corresponding momentum and energy conservation laws, which contain explicit contributions from the pair distribution function. 
We showed that, in thermal equilibrium, solutions with uniform and time-independent number density and specific entropy, along with the absence of macroscopic motion, satisfy the equations of motion. 
Moreover, the equilibrium energy in this case reproduces the standard thermodynamic result for interacting systems. 
Expanding the equations to linear order in the displacement field around these equilibrium solutions, transverse and longitudinal dispersion relations were obtained. 
The transverse dispersion law coincides exactly with the QLCA predictions. 
In contrast, the longitudinal mode contains additional terms that account for thermal effects arising from both the microscopic kinetic energy and the correlations captured by derivatives of the pair distribution function.
\par

The only quantities required for numerical evaluation of the dispersion laws are the adiabatic derivative of the temperature and the pair distribution function, along with its derivatives. 
To compute the adiabatic derivative, knowledge of both the energy and pressure in thermal equilibrium is required.
While accurate fits for the excess internal energy that are valid across a wide range of coupling and screening parameters are available~\cite{PhysRevE.66.016404}, similarly reliable fits for the excess pressure are missing. 
To address this, we computed the pressure from the Helmholtz free energy, which was obtained from thermodynamic integration. 
For the pair distribution function, we employed a step function approximation~\cite{10.1063/5.0194352, 10.1063/1.4942169, 10.1063/1.5088141}, which greatly simplifies the otherwise complex integrals appearing in the dispersion law, enabling fast and efficient computations. 
The cutoff radius was determined by requiring the internal energy, computed using the approximate pair distribution function, to be consistent with the values in thermal equilibrium. 
We examined the longitudinal dispersion relation for values $0.5 \leq \kappa_0 \leq 3$ and demonstrated that the typical hydrodynamic behavior, where the frequency increases monotonically with the wavenumber, is modified as the coupling parameter approaches its melting value. 
Specifically, in the range $0 \leq |\vec{q}_0| \leq 5$, the dispersion relation at strong coupling develops a local maximum, whose position shifts to higher wavenumbers as $\kappa_0$ increases. 
Following this maximum, a local minimum appears, the position of which depends only weakly on $\kappa_0$.
\par 

To assess the accuracy in the long-wavelength limit, we examined the speed of sound.
When compared to MD data~\cite{PhysRevE.100.063206}, our variational approach shows excellent agreement for $\kappa_0 = 0.5$ and $\kappa_0 = 1$, with maximum errors below $3\%$ across the entire range of coupling strengths considered. At these screening values, the deviations are comparable to those from the standard hydrodynamic approach, which augments the Euler equations with a mean-field term. However, the errors increase for stronger screening, $\kappa_0=2$ and $\kappa_0=3$. Here, the largest discrepancies occur at relatively low coupling, when $\Gamma$ is a few percent of the melting value. As the coupling strength increases, the errors decrease, and our approach converges first to the extended QLCA, and at even higher coupling, to the standard QLCA. Meanwhile, the mean-field hydrodynamic approach continues to match simulation results well even at strong screening.
\par 

To assess the accuracy at short length scales, we examined the dispersion relation at finite wavenumbers for $\kappa_0 = 1$ and $\kappa_0 = 2$.
When compared to the peaks in the longitudinal current fluctuation spectrum obtained from our simulations, we find that, for both screening values and across a wide range of coupling parameters, our variational approach reasonably reproduces the data. 
At weak coupling, our model outperforms other theoretical approaches. 
At moderate coupling, it slightly overestimates the dispersion and shows convergence toward the extended QLCA results, while QLCA aligns best with the data. 
At strong coupling, our theory converges to both QLCA and extended QLCA. 
Considering the full coupling range, our approach provides the best overall agreement with the MD data compared to the other theoretical models tested. 
This demonstrates the ability of our theory to make accurate predictions at short length scales, well beyond the typical applicability of standard hydrodynamic descriptions.
\par 

We verified through our own simulations that the discrepancies observed for the sound speed at higher screening values cannot be attributed to inaccuracies in the equation of state or to the step function approximation of the pair distribution function.
Note that our current framework does not include the effects of heat transport, relaxation, or dissipation.
Although the mean-field approach likewise neglects heat conduction and dissipative processes such as viscosity, our model assumes a particular form for the out-of-equilibrium behavior of the pair distribution function, implying a specific relaxation toward local thermodynamic equilibrium. 
It is possible that at higher screening, relaxation behaves differently, and a modified ansatz for the pair distribution function may be required, for example, as in the Eulerian variational principle discussed for the OCP in the original formulation of the considered variational framework \cite{10.1063/5.0194352}.
Additionally, one could compute the isothermal speed of sound, as opposed to the isentropic (adiabatic) case considered in this work, which would correspond to the opposite limit, namely, infinite heat conductivity.
This could provide insight into the role of thermal conduction without explicitly introducing a separate heat transport equation.
\par 

In summary, we have presented hydrodynamic equations that provide a consistent description of the YOCP and can be used to solve arbitrary linear and nonlinear problems of interest.  
Moreover, because the variational principle and the resulting equations are formulated for an arbitrary interaction potential, they can readily be applied to other strongly coupled systems with different interactions. 
The successful treatment of the OCP in earlier work and of the YOCP in this paper suggests that such extensions will again yield accurate and efficient descriptions.
\par 

\begin{acknowledgments}
The authors thank S. Putterman and M. Bonitz for many valuable discussions.
This research has been funded by the German Science Foundation (DFG) via grant BO1366-13/2.
The authors gratefully acknowledge the computing time made available to them on the high-performance computer Lise at the NHR Center NHR@ZIB. This center is jointly supported by the Federal Ministry of Education and Research and the state governments participating in the National High-Performance Computing (NHR) joint funding program (http://www.nhr-verein.de/en/our-partners).
\end{acknowledgments}

\appendix
\section{Energy conservation law}
\label{appendix:energy_law}

The energy conservation law is given by 
\begin{equation}
	\label{eq:energy_conservation_law}
	\begin{gathered}
		\frac{ \partial \mathcal{E}}{\partial t} + \sum \limits_{j=1}^3 \frac{\partial J_j}{\partial a_j} = \sigma,
	\end{gathered}
\end{equation}
with the energy density defined by
\begin{equation}
	\label{eq:energy_density}
	\begin{gathered}
		\mathcal{E} = \frac{1}{2}mn^{\mathrm{ref}} \left( \frac{\partial \vec{x}}{\partial t} \right)^2 + \frac{3}{2} n^{\mathrm{ref}} k_B T + \frac{1}{2} \int n^{\mathrm{ref}} (n^{\mathrm{ref}})^T \phi  \left( \frac{g + g^T}{2} \right) \mathrm{d}\vec{a}',
	\end{gathered}
\end{equation}
the energy flux
\begin{equation}
	\label{eq:energy_flux}
	\begin{gathered}
		J_j = - \sum \limits_{i=1}^3 \frac{\partial x_i}{\partial t} \frac{\partial (\mathrm{det}(\partial \vec{x}/\partial \vec{a}))}{\partial (\partial_j x_i)}  \frac{\partial T}{\partial (1/n)} \bigg|_s \left\{ \frac{3}{2} k_B + \frac{1}{2} \int (n^{\mathrm{ref}})^T \phi \frac{\partial g}{\partial T} \bigg|_{n^{\mathrm{ref}}, |\vec{a}-\vec{a}'|} \mathrm{d}\vec{a}' \right\},
	\end{gathered}
\end{equation}
and the nonlocal source term
\begin{equation}
	\label{eq:energy_nonlocal_source}
	\begin{gathered}
		\sigma = \sum \limits_{i=1}^3 \frac{1}{2} \int n^{\mathrm{ref}} (n^{\mathrm{ref}})^T \left( \frac{g + g^T}{2} \right) \left( \frac{\partial \phi}{\partial x_i^T}\bigg|_{\vec{x}} \frac{\partial x_i^T}{\partial t} - \frac{\partial \phi}{\partial x_i}\bigg|_{\vec{x}^T} \frac{\partial x_i}{\partial t} \right) \mathrm{d}\vec{a}'
		\\
		+ \sum \limits_{i=1}^3 \sum \limits_{j=1}^3 \frac{1}{2} \int \phi \bigg( n^{\mathrm{ref}} \frac{\partial \left(g^T/2\right)}{\partial T^T} \bigg|_{(n^{\mathrm{ref}})^T, |\vec{a}-\vec{a}'|} \frac{\partial T^T}{\partial (1/n^T)} \bigg|_{s^T} \frac{\partial (\mathrm{det}(\partial \vec{x}^T/\partial \vec{a}'))}{\partial (\partial'_j x^T_i)} \frac{\partial^2 x_i^T}{\partial t \partial a'_j} 
		\\
		-(n^{\mathrm{ref}})^T  \frac{\partial \left(g/2\right)}{\partial T} \bigg|_{n^{\mathrm{ref}}, |\vec{a}-\vec{a}'|} \frac{\partial T}{\partial (1/n)} \bigg|_s \frac{\partial (\mathrm{det}(\partial \vec{x}/\partial \vec{a}))}{\partial (\partial_j x_i)} \frac{\partial^2 x_i}{\partial t \partial a_j}  \bigg) \mathrm{d}\vec{a}'.
	\end{gathered}
\end{equation}
\par

\section{Linearized equations of motion}
\label{appendix:linearized_eom}

Using the solution form given in Eq.~\eqref{eq:linearized_x}, the equations of motion from Eq.~\eqref{eq:equations_of_motion} yield the first-order correction
\begin{equation}
	\begin{gathered}
		m \frac{\partial^2 {\xi}_i}{\partial t^2} = - \frac{\partial (\vec{\nabla} \cdot \vec{\xi})}{\partial a_i} \bigg( \frac{\partial}{\partial n_0} \bigg|_{s_0} \left( \frac{3}{2} k_B \frac{\partial T_0}{\partial (1/n_0)} \bigg|_{s_0}  \right) \\
		+ \frac{n_0}{2} \frac{\partial^2 T_0}{\partial n_0 \partial (1/n_0)} \bigg|_{s_0} \int \phi (|\vec{r}|) \frac{\partial g_0}{\partial T_0} \bigg|_{n_0, |\vec{r}|} \mathrm{d}\vec{r} \\
		+\frac{n_0}{2} \frac{\partial T_0}{\partial (1/n_0)} \bigg|_{s_0} \frac{\partial T_0}{\partial n_0} \bigg|_{s_0} \int \phi (|\vec{r}|) \frac{\partial^2 g_0}{\partial T_0^2} \bigg|_{n_0, |\vec{r}|} \mathrm{d}\vec{r} \bigg) \\
		+ \frac{1}{2} \frac{\partial T_0}{\partial (1/n_0)} \bigg|_{s_0} \sum \limits_{j=1}^3 \frac{\partial}{\partial a_i} \int \frac{\partial \phi}{\partial a_j}\bigg|_{\vec{a}'} \frac{\partial g_0}{\partial T_0} \bigg|_{n_0, |\vec{a}-\vec{a}'|} \left( \xi_j - \xi^T_j \right) \mathrm{d}\vec{a}' \\
		+ \frac{n_0^2}{2} \frac{\partial T_0}{\partial n_0} \bigg|_{s_0} \int \frac{\partial \phi}{\partial a_i}\bigg|_{\vec{a}'}\frac{\partial g_0}{\partial T_0} \bigg|_{n_0, |\vec{a}-\vec{a}'|} \left( \vec{\nabla} \cdot \vec{\xi} + \vec{\nabla}' \cdot \vec{\xi^T} \right) \mathrm{d}\vec{a}' \\
		- \sum \limits_{j=1}^3 n_0 \int \frac{\partial^2 \phi}{\partial a_i \partial a_j}\bigg|_{\vec{a}'} g_0\left(n_0, T_0, |\vec{a}-\vec{a}'| \right) \left( \xi_j - \xi^T_j \right) \mathrm{d}\vec{a}'.
	\end{gathered}
\end{equation}
As before, the subscript $``0"$ indicates that a function is evaluated at equilibrium.
\par 

For the Fourier transform we use the convention
\begin{equation}
	\begin{gathered}
		\vec{\Psi}(\vec{k}, t) = \widehat{\vec{\xi}}(\vec{k}, t) = \int \vec{\xi}(\vec{a}, t) e^{-i\vec{k}\cdot \vec{a}} \mathrm{d}\vec{a},
	\end{gathered}
\end{equation}
in which case, for any functions $f$ and $g$ that depend on $\vec{a}$, we have $\widehat{\partial f/\partial a_j} = ik_j \widehat{f}$ and $\widehat{f \star g} = \widehat{f}\widehat{g}$, where $\star$ denotes convolution.
In addition, we use the identities 
\begin{equation}
	\label{eq:yukawa_derivatives}
	\begin{gathered}
		\frac{\partial \phi}{\partial r_i} = -\frac{q^2}{4 \pi \varepsilon_0} \frac{r_i e^{-|\vec{r}|/\lambda}}{|\vec{r}|^3} \left(1 + \frac{|\vec{r}|}{\lambda} \right), \\
		\frac{\partial^2 \phi}{\partial r_i \partial r_j} = \frac{q^2}{4 \pi \varepsilon_0} \frac{e^{-|\vec{r}|/\lambda}}{|\vec{r}|^5} \left( r_i r_j \left( 3 + 3\frac{|\vec{r}|}{\lambda} + \frac{|\vec{r}|^2}{\lambda^2} \right) - \delta_{ij} |\vec{r}|^2 \left( 1 + \frac{|\vec{r}|}{\lambda} \right) \right) - \frac{q^2}{3 \varepsilon_0} \delta_{ij} \delta(\vec{r}),
	\end{gathered}
\end{equation}
derived by delta-function properties \cite{10.1119/1.13127}.
\par 

Combining these results leads to 
\begin{equation}
	\begin{gathered}
		m \frac{\partial^2 \vec{\Psi}}{\partial t^2} = \vec{k} (\vec{k} \cdot \vec{\Psi}) \bigg( \frac{\partial}{\partial n_0} \bigg|_{s_0} \left( \frac{3}{2} k_B \frac{\partial T_0}{\partial (1/n_0)} \bigg|_{s_0}  \right) \\
		+ \frac{q^2 n_0}{8 \pi \varepsilon_0} \frac{\partial^2 T_0}{\partial n_0 \partial (1/n_0)} \bigg|_{s_0} \int \frac{e^{-|\vec{r}|/\lambda}}{|\vec{r}|} \frac{\partial g_0}{\partial T_0} \bigg|_{n_0, |\vec{r}|} \mathrm{d}\vec{r} \\
		+\frac{q^2 n_0}{8 \pi \varepsilon_0} \frac{\partial T_0}{\partial (1/n_0)} \bigg|_{s_0} \frac{\partial T_0}{\partial n_0} \bigg|_{s_0} \int \frac{e^{-|\vec{r}|/\lambda}}{|\vec{r}|} \frac{\partial^2 g_0}{\partial T_0^2} \bigg|_{n_0, |\vec{r}|} \mathrm{d}\vec{r} \bigg) \\
		- \frac{iq^2 \vec{k}}{8 \pi \varepsilon_0} \frac{\partial T_0}{\partial (1/n_0)} \bigg|_{s_0} \int \frac{(\vec{r} \cdot \vec{\Psi}) e^{-|\vec{r}|/\lambda}}{|\vec{r}|^3} \left(1 + \frac{|\vec{r}|}{\lambda} \right) \frac{\partial g_0}{\partial T_0} \bigg|_{n_0, |\vec{r}|} \left( 1 - e^{-i\vec{k} \cdot \vec{r}} \right) \mathrm{d}\vec{r} \\
		+ \frac{iq^2 (\vec{k} \cdot \vec{\Psi})}{8 \pi \varepsilon_0 }\frac{\partial T_0}{\partial (1/n_0)} \bigg|_{s_0} \int \frac{\vec{r} e^{-|\vec{r}|/\lambda}}{|\vec{r}|^3} \left(1 + \frac{|\vec{r}|}{\lambda} \right) \frac{\partial g_0}{\partial T_0} \bigg|_{n_0, |\vec{r}|} \left( 1 + e^{-i\vec{k} \cdot \vec{r}} \right) \mathrm{d}\vec{r} \\
		- \frac{q^2 n_0}{4 \pi \varepsilon_0} \int g_0 e^{-|\vec{r}|/\lambda} \bigg( \frac{\vec{r} (\vec{r} \cdot \vec{\Psi})}{|\vec{r}|^5} \left( 3 + 3\frac{|\vec{r}|}{\lambda} + \frac{|\vec{r}|^2}{\lambda^2} \right) \\
		- \frac{\vec{\Psi}}{|\vec{r}|^3} \left( 1 + \frac{|\vec{r}|}{\lambda} \right)  \bigg)  \left( 1 - e^{-i\vec{k} \cdot \vec{r}} \right) \mathrm{d}\vec{r},
	\end{gathered}
\end{equation}
which is then used to obtain Eqs.~\eqref{eq:general_transverse_law} and \eqref{eq:general_longitudinal_law}.

\bibliography{Yukawa_OCP_bibliography}

\end{document}